  \UseRawInputEncoding
\documentclass[journal]{IEEEtran}

\usepackage[colorlinks,linkcolor=blue,anchorcolor=gray,citecolor=blue,urlcolor=black]{hyperref}

\usepackage{amsmath,amsfonts,amssymb}
\usepackage{algorithm}
\usepackage{algorithmicx}
\usepackage{booktabs}
\usepackage{multirow}
\usepackage{longtable}
\usepackage{graphicx}
\usepackage{epstopdf}
\usepackage[square, comma, sort&compress, numbers]{natbib} %when using several ref. together
\usepackage[numbers]{natbib}
\usepackage{cleveref}
\crefname{table}{Tab.}{Tabs.}
\crefname{figure}{Fig.}{Figs.}
 \usepackage{url}

\newtheorem{claim}{Claim}

\ifCLASSINFOpdf
\else
\fi
\begin{document}
%
% paper title
% Titles are generally capitalized except for words such as a, an, and, as,
% at, but, by, for, in, nor, of, on, or, the, to and up, which are usually
% not capitalized unless they are the first or last word of the title.
% Linebreaks \\ can be used within to get better formatting as desired.
% Do not put math or special symbols in the title.
\title{Fast hyperspectral image denoising and inpainting\\based on low-rank and sparse representations}
%
%
% author names and IEEE memberships
% note positions of commas and nonbreaking spaces ( ~ ) LaTeX will not break
% a structure at a ~ so this keeps an author's name from being broken across
% two lines.
% use \thanks{} to gain access to the first footnote area
% a separate \thanks must be used for each paragraph as LaTeX2e's \thanks
% was not built to handle multiple paragraphs
%

\author{Lina~Zhuang,~\IEEEmembership{Student Member,~IEEE,}
        and~Jos\'{e}~M.~Bioucas-Dias,~\IEEEmembership{Fellow,~IEEE}
       % <-this % stops a space
\thanks{
Manuscript received July 21, 2017; revised October 24, 2017; accepted
January 7, 2018. Date of publication February 11, 2018; date of current
version March 9, 2018. 
This work was supported in part by the  European Union's Seventh Framework Programme (FP7-PEOPLE-2013-ITN) under grant agreement n$^{\text o}$ 607290 SpaRTaN
and in part by the  Funda\c{c}\~{a}o para a Ci\^encia e Tecnologia, Portuguese Ministry of Science and Higher Education, projects UID/EEA/50008/2013 and ERANETMED/0001/2014.}% <-this % stops a space
\thanks{
The authors are with  the Instituto de Telecomunica\c{c}\~{o}es, Instituto Superior T\'{e}cnico, Universidade de Lisboa, 1049-001, Lisbon, Portugal (e-mail: lina.zhuang@lx.it.pt; bioucas@lx.it.pt).}
\thanks{
Digital Object Identifier 10.1109/JSTARS.2018.2796570}
}

% The paper headers
%\markboth{IEEE JOURNAL OF SELECTED TOPICS IN APPLIED EARTH OBSERVATIONS AND REMOTE SENSING, VOL. 11, NO. 3, MARCH 2018}%
%{Shell \MakeLowercase{\textit{et al.}}: Bare Demo of IEEEtran.cls for IEEE Journals}
% The only time the second header will appear is for the odd numbered pages
% after the title page when using the twoside option.
% 
% *** Note that you probably will NOT want to include the author's ***
% *** name in the headers of peer review papers.                   ***
% You can use \ifCLASSOPTIONpeerreview for conditional compilation here if
% you desire.

% If you want to put a publisher's ID mark on the page you can do it like
% this:
%\IEEEpubid{0000--0000/00\$00.00~\copyright~2015 IEEE}
% Remember, if you use this you must call \IEEEpubidadjcol in the second
% column for its text to clear the IEEEpubid mark.

% use for special paper notices
%\IEEEspecialpapernotice{(Invited Paper)}

% make the title area
\maketitle

% As a general rule, do not put math, special symbols or citations
% in the abstract or keywords.
\begin{abstract}
This paper introduces two very fast and competitive hyperspectral image (HSI) restoration algorithms: {\em Fast Hyperspectral Denoising} (FastHyDe), a denoising algorithm able to cope with Gaussian and Poissonian noise, and   {\em Fast Hyperspectral Inpainting} (FastHyIn), an inpainting algorithm to restore HSIs where some observations from known pixels in some known bands are missing.  FastHyDe and FastHyIn fully exploit  extremely compact and sparse HSI representations  linked with their low-rank  and self-similarity characteristics.  In a series of experiments with simulated and real data, the newly introduced  FastHyDe and FastHyIn compete with state-of-the-art methods, with much  lower computational complexity. \textcolor{red}{A MATLAB demo of this work is available
at \url{www.lx.it.pt/~bioucas/code/Demo_FastHyDe_FastHyIn.rar}  for the
sake of reproducibility.}
\end{abstract}

% Note that keywords are not normally used for peerreview papers.
\begin{IEEEkeywords}
High dimensional data, low dimensional subspace, non-local patch(cube), self-similarity, BM3D, BM4D, low-rank regularized collaborative filtering. 
\end{IEEEkeywords}

% For peer review papers, you can put extra information on the cover
% page as needed:
% \ifCLASSOPTIONpeerreview
% \begin{center} \bfseries EDICS Category: 3-BBND \end{center}
% \fi
%
% For peerreview papers, this IEEEtran command inserts a page break and
% creates the second title. It will be ignored for other modes.
\IEEEpeerreviewmaketitle

\section{Introduction}

Hyperspectral remote sensing images have been widely used in countless applications (e.g., earth observation, environmental protection and natural disaster monitoring), owing to their remarkably high spectral resolution (hundreds or thousands spectral channels), which enables precise material identification  via spectroscopic analysis \cite{overview}. However,  this potential is 
often compromised due to low quality of the hyperspectral images (HSIs), linked with various  degradation mechanisms, such as electronic noise, Poissonian noise, quantization noise, stripe noise, and atmospheric effects. 

As the number of spectral bands increases in the new-generation hyperspectral sensors, the spectral bandwidth decreases implying that, everything else kept constant, each spectral channel receives less photons, yielding higher levels of Poissonian noise. So Poissonian noise is becoming the main concern in real hyperspectral images \cite{poissonRef1,poissonRef2,poisson_Qian}. If the mean values of the photon counts is larger than 4,  then Poissonian noise can be converted into approximately additive Gaussian noise with nearly constant variance using variance-stabilizing transformations \cite{VarianceStabilizing}. This opens a door to use available algorithms designed for additive Gaussian noise \cite{FastHyDe}. In this paper, we assume  that the observation noise is either additive Gaussian or Poissonian. In the latter case, and prior to restore the noisy image, the Poissonian noise is converted into approximate additive Gaussian noise by applying a variance-stabilizing transformation.

Natural images are self-similar. This means that they  contain many similar patches  at different locations and scales. This characteristic   has been recently
exploited by the patch-based image restoration methods and  holds the state-of-the-art in image denoising. Representative examples of this methodology in single-band images include the non-local means filter \cite{buades2005non}, the Gaussian mixture model (GMM) learned from the noisy image \cite{teodoro2015single}, and the collaborative filtering of groups  of similar patches BM3D \cite{BM3D}, LRCF \cite{LRCF}, and  EPLL \cite{zoran2011learning}.
Identical ideas have been pursued in multi-band image denoising:
BM4D \cite{BM4D}, VBM4D \cite{VBM4D}, and MSPCA-BM3D \cite{MSPCA-BM3D} 
use collaborative filtering in groups of 3D patches extracted from volumetric data, videos, multispectral data, respectively.
DHOSVD \cite{rajwade2013image_HOSVD} applies hard  threshold filtering to the   higher order SVD coefficients of similar patches.

In HSI denoising, besides  the spatial information, the high correlation in spectral domain has also been widely investigated, as it implies that the spectral vectors live in low-dimensional manifolds or subspaces \cite{overview, Hysime}, admitting, therefore, extremely compact and sparse representations on suitable frames.  These characteristics have been exploited, namely, in low-rank matrix approximation \cite{NAILRMA,LR_zhang1,LR_zhang2,LR_zhang3,7180362}, in the structured tensor TV-based regularization \cite{SSAHTV}, and in the adaptive spectrum-weighted sparse Bayesian dictionary learning method (ABPFA) \cite{CS_spectra}.   Sparse representations have also been exploited in many inverse problems in hyperspectral imaging, for example, classification (see more details in \cite{fang2014spectral,ShutaoLiref2,ShutaoLiref3}).

Image inpainting is the process of image reconstruction from incomplete observation.  For single-band images, spatial information from areas with complete observations is used to inpaint the incomplete observations, i.e., the missing pixels.
Inpainting techniques mainly exploit spatial information of consistent geometric structure \cite{Bertalmio2002Image}, texture with repetitive patterns \cite{efros1999texture}, and  combinations of both \cite{textureSynthesis}.  
Structural inpainting focuses on the consistency of the geometric structure by enforcing a smoothness prior to preserve edges or isophotes. The consistency of structure may be enforced using  partial differential equations \cite{PDE}, total variation \cite{chan2006total,SSAHTV}, and  Markov random field image priors \cite{shen2009map}. But enforcing consistent structure does not restore texture. This calls for textural inpainting, which tries to capture textures with a repetitive pattern and complete the missing region using its similar neighbourhood. Textural inpainting may not be able to keep consistency in the boundaries between image regions.  Since most part of an image consist of structure and texture, the state-of-the-art inpainting methods attempt to combine structural and textural inpainting. For example, \cite{textureSynthesis} decomposes the image into the sum of two functions with different basic characteristics, and then reconstruct each one of these functions separately with structure and texture inpainting algorithms. 
Exemplar-based image inpainting fills missing regions of an image by searching for similar patches in a neighboring region of the image, and copying the pixels from the most similar patch into the missing region \cite{criminisi2004region,buades2005non}.

For multispectral and HSIs inpainting, coherence lying in temporal, spatial, or spectral domains is exploited. For example, the spatiotemporal relationships between a sequence of multitemporal images were studied in \cite{temporal1,temporal2} to reconstruct corrupted regions. Spectral coherence between a damaged band and other bands was exploited in \cite{CS_spectra}. In \cite{liangpei_inp}, both spatial and spectral coherence were achieved via a multichannel non-local total variation inpainting model.

From the above considerations, it may be concluded that denoising and inpainting approaches share a number of similar techniques to promote spatial-spectral coherence. In fact, from the formal point of view, the inpainting restoration problem may be interpreted as a denoising one with missing observations, whose locations are known
 \cite{liangpei_inp}.

\subsection{Contribution}

Most of the published hyperspectral denoising and inpainting algorithms are time-consuming, namely due to the large
sizes of HSIs  and, often, due to the implementation of iterative  estimation procedures both in the spatial and the spectral domains. 
In an effort to mitigate these shortcomings, this paper  introduces a  fast hyperspectral denoising (FastHyDe) approach and its   fast hyperspectral inpainting  (FastHyIn) extension. 

FastHyDe and FastHyIn take full advantage of  the HSIs' low-rank structure and  self-similarity  referred to above: 1)~the low-rank structure of HSI is fully exploited by  representing the spectral vectors of the clean image in a pre-learned subspace; 2)~we claim that the images of subspace coefficients  are self-similar (see Claim \ref{claim}) and can be denoised component by component using non-local patch-based single band denoisers. Groups of similar non-local  patches admit sparse representations which underlies strong noise attenuation.

This work is an extension of the material published in \cite{FastHyDe}.  The new material is the following: a) FastHyDe is herein introduced and characterized in more detail; b) a new algorithm, termed FastHyIn,  to solve inpainting inverse problems is  introduced; and c) exhaustive array of experiments and comparisons is carried out.

The paper is organized as follows. Section~\ref{sec:fasthyde} introduces formally FastHyDe, a denoising approach based on low-rank and sparse representations. Its inpainting version, FastHyIn, is given in Section~\ref{sec:fasthyin}. Section \ref{sec:exp} and \ref{sec:exp_real}  present experimental results including comparisons with the state of the art. Section~\ref{sec:concl} concludes the paper.

\label{sec:intro}

\section{Formulation and proposed denoiser}

\label{sec:fasthyde}

Let ${\bf X} :=[{\bf x}_1,\dots{\bf x}_n]\in \mathbb{R}^{n_b\times
n}$ denote a  HSI with $n$ spectral vectors (the columns of $\bf X$) of size $n_b$. The rows $\bf X$ contain  $n_b$ spectral bands, which are images, corresponding to the scene reflectance in a given wavelength interval,  with $n$ pixels organized in a grid in the spatial domain.
In hyperspectral denoising problems under the additive noise assumption, the observation model may be written as
\begin{equation}
\label{obs}
\mathbf{Y = X + N},
\end{equation}
where $\mathbf{Y,N} \in \mathbb{R}^{n_b \times n}$ represent the observed HSI data and noise, respectively.  

We assume that the spectral vectors ${\bf x}_i$, for $i=1,\dots,n$, live in a $k-$dimensional subspace ${\cal S}_k$, with $n_b\gg k$. This is a very good approximation in most real HSIs \cite{overview}. Therefore, we may write
\begin{equation}
\mathbf{X=EZ},
\end{equation}
where the columns of $\mathbf{E} =[{\bf e}_1,\dots,{\bf e}_k] \in \mathbb{R}^{n_b \times k}$  holds a basis 
for ${\cal S}_k$ and matrix $\mathbf{Z} \in \mathbb{R}^{k \times n}$ holds the representation  coefficients of $\bf X$ with respect to (w.r.t.) $\bf E$. We assume,
without loss of generality, that $\bf E$ is semi-unitary, that is ${\bf E}^T {\bf E} = {\bf I}_k$ with ${\bf I}_k$ representing the identity matrix of dimension $k$.
Matrix  $\bf E$ may be learned from the data using, e.g.,  the HySime algorithm \cite{Hysime} or  singular value decomposition (SVD) of $\bf Y$ in the case the noise is independent and identically distributed (i.i.d.). We will herein term  the images associated with the rows of $\bf Z$ as \textit{eigen-images}. 

Images of the real world are self-similar, that is, they contain many similar patches at different locations and scales. The exploitation of self-similarity, as a form of prior knowledge/regularization, underlies the state-of-the-art in imaging inverse problems. 
HSI bands, being images of the real world, are self-similar. Furthermore, since each band corresponds to the reflectance, in a given wavelength interval, of the same surface, the spatial structure of the self-similarity is identical across all bands. This has a fundamental relevance for the proposed approach stated in the following claim.

\begin{claim}
\label{claim}
	The eigen-images associated to the rows of  ${\bf Z}$   are self-similar. 
\end{claim}

\vspace{-0.1cm}
\noindent
{\em Justification:}  Since $\bf E$ is semi-unitary, then ${\bf Z}= {\bf E}^T{\bf X}$, and therefore the eigen-images are linear combinations of the the bands of $\bf X$. Given that the bands of $\bf X$ are self-similar and have the same  self-similarity  spatial structure, then  the eigen-images are self-similar.\hfill $\square$

\vspace{0.3cm}

\begin{figure}[htbp]
\centering
\includegraphics[scale=0.5]{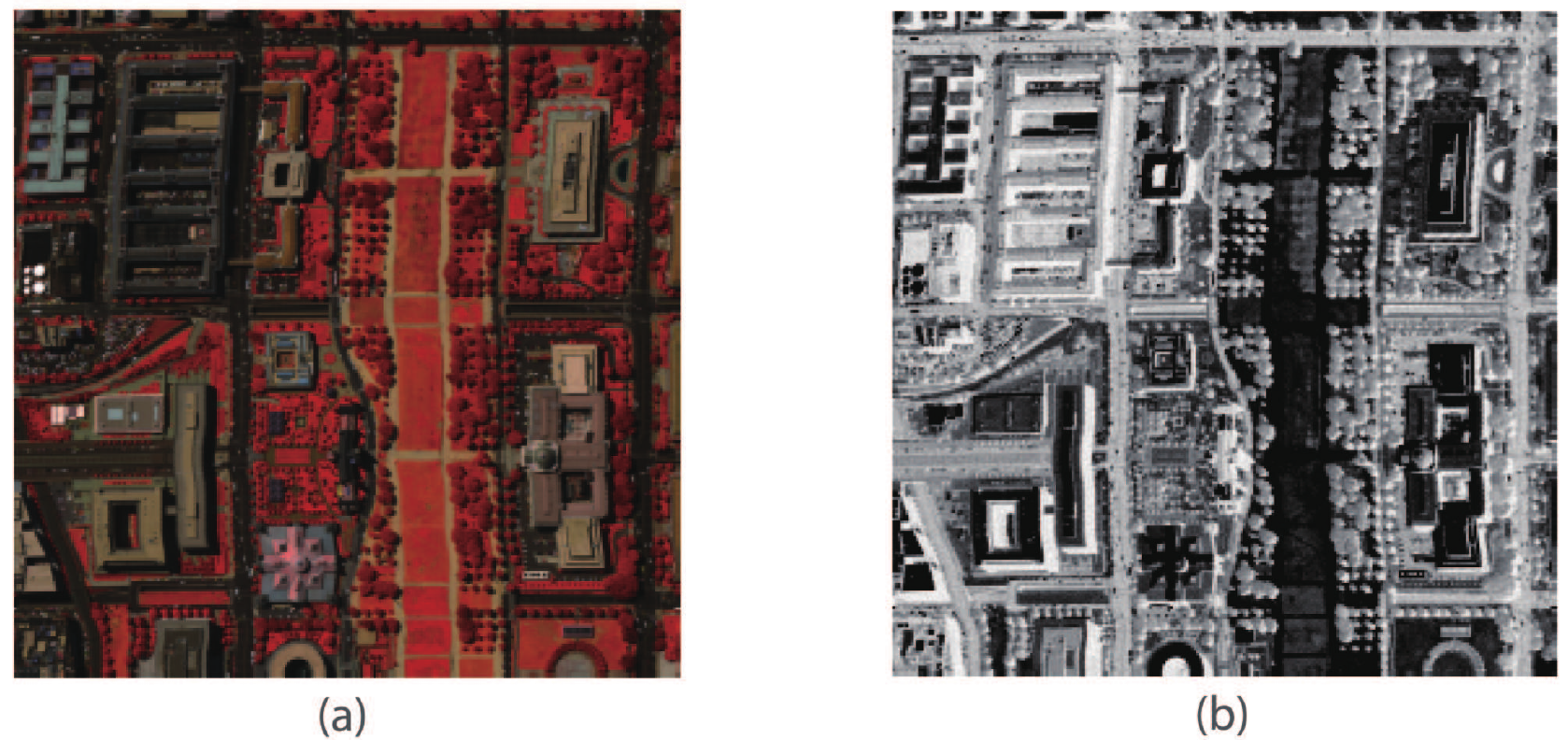}
\caption{False color composites of the Washington DC Mall subscene (a) and its 1st band of eigen-images (b).}
\label{fig:img_eigen}
\end{figure}

Fig. \ref{fig:img_eigen} shows in the left hand side a false color image of the a Washington DC Mall subscene and in the right hand side the eigen-image corresponding to the first column of $\bf E$. Clearly, the eigen-image preserves the self-similar structure of the corresponding  color image.

This paper explores two main characteristics of  hyperspectral data: 1) HSIs live in low dimensional subspaces, which opens the door to remove the bulk of
the noise using projection-based methods \cite{overview}; 2) the eigen-images are self-similar and, therefore, they may be denoised with non-local patch-based methods such
as BM3D \cite{BM3D} and LRCF \cite{LRCF}.

Below we start by considering that  the noise is additive Gaussian and i.i.d. Later, we consider  additive Gaussian non-i.i.d and Poissonian noise.

\subsection{Additive Gaussian i.i.d. noise}

\label{sec:iid}
In the first case, we consider that the noise is additive zero-mean Gaussian and i.i.d. over all components of $\bf N$. 
Assuming that the subspace $\mathbf{E}$ has been learned from the observed data $\bf Y$, the eigen-images denoising problem is formulated as 
\begin{equation}
\label{equ:iid}
\begin{split}
\widehat{\mathbf{Z}}& \in  \arg\min_{\mathbf{Z}} \frac{1}{2}\Vert \mathbf{EZ-Y} \Vert_F^2 +\lambda \phi (\mathbf{Z})\\
 &= \arg\min_{\mathbf{Z}} \frac{1}{2}\Vert \mathbf{Z-E}^T{\bf Y} \Vert_F^2 +\lambda \phi (\mathbf{Z}),
\end{split}
\end{equation}
where $\|{\bf X}\|_F = \sqrt{\text{trace}({\bf XX}^T)}$  is the Frobenius norm of 
$\bf X$. 
The optimization problems in \eqref{equ:iid}  are equivalent since ${\bf E}$ is semi-unitary and ${\bf E}^T {\bf E} = {\bf I}$.
The first term on the right-hand of side \eqref{equ:iid} represents the data fidelity
and accounts for the zero-mean Gaussian i.i.d noise,  while the second term is a regularizer expressing  prior information tailored to self-similar images. 
 
We assume that the function $\phi$ is decoupled w.r.t. the eigen-images, that is 
\begin{equation}
\label{eq:phi_decoupled}
\phi(\mathbf{Z}) = \sum_{i=1}^k \phi_i(\mathbf{Z}^i)
\end{equation}
where $\mathbf{Z}^i$ is the $i$th  eigen-image, i.e., the $i$th row of $\bf Z$. An informal justification for the decoupling  is that the components of $\bf Z$ tend to be decorrelated. Although decorrelation does not imply statistical independence,  it is  a necessary condition for it. In practice, this  assumption   leads to excellent results as shown in Section \ref{sec:exp}. Another relevant observation is that the noise associated to the term ${\bf E}^T\bf Y$ is still i.i.d.  with the same variance as the original  components of  $\bf N$. In fact, 
$$
 {\bf E}^T{\bf Y}= {\bf Z} + {\bf E}^T{\bf N},
$$
and therefore the columns of the noise  term ${\bf E}^T{\bf N}$ are independent and, given $\bf n$, a generic column of $\bf N$, we have 
$$
 \mathbb{E}[{\bf E}^T{\bf n}{\bf n}^T{\bf E}] 
    = {\bf E}^T\mathbb{E}[{\bf n}{\bf n}^T]{\bf E}
    = \sigma^2{\bf E}^T{\bf E}
    =\sigma^2{\bf I}_k,
$$
where $\sigma^2$ is the variance of a generic element of $\bf N$. Therefore, ${\bf E}^T{\bf n} \sim {\cal N}({{\bf 0}, \sigma^2{\bf I}_k})$, this is, ${\bf E}^T{\bf n}$ is zero-mean Gaussian with covariance $\sigma^2{\bf I}_k$.

Under the hypothesis \eqref{eq:phi_decoupled}, the solution of (\ref{equ:iid})
is decoupled w.r.t. ${\bf Z}^i$ and may be written as 
\begin{equation}
\widehat{\mathbf{Z}} = \psi_{\lambda \phi}(\mathbf{E}^T {\bf Y})=\left[ \begin{array}{c}
\psi_{{\lambda \phi}_1}(\mathbf{e}_1^T\mathbf{Y}) \\
\vdots\\
\psi_{{\lambda \phi}_k}(\mathbf{e}_k^T\mathbf{Y})
\end{array} \right],
\end{equation}
where
\begin{equation}
\label{eq:denoising_z}
 \psi_{{\lambda \phi}_i}({\bf y}) = \arg\min_{\bf w}
   \frac{1}{2}\|{\bf y-w}\|^2_F +\lambda\phi_i({\bf w})
\end{equation}
is the so-called denoising operator, or Moreau proximity operator of $\phi_i$ \cite{moreau_Combettes}. In this paper, we resort to the plug-and-play (PnP) prior framework \cite{PnP_Venkatakrishnan} to address  the denoising problem (\ref{eq:denoising_z}).
The central idea in PnP is, instead of investing efforts in tailoring regularizers  promoting self-similar images and then computing its proximity operators, to use directly  a state-of-the-art denoiser conceived to enforce self-similarity, such as BM3D \cite{BM3D}, LRCF \cite{LRCF}, and GMM \cite{teodoro2015single}. A pertinent question in the PnP approach is that, given a denoiser, weather there exists a convex regularizer of which the denoiser is the proximity operator. The answer for BM3D, LRCF, and GMM is negative \cite{teodoro2017sharpening,ljubenovic2017blind}, as it is for most state-of-the-art denoisers. This fact should not prevent us, however, to use such denoisers as they are effective in promoting self-similar images. In this work, we  selected BM3D, as it is the state-of-the-art and a very fast implementation thereof is publicly available.

The proposed algorithm for HSI denoising, termed {\em Fast Hyperspectral denoising} (FastHyDe), is summarized in Algorithm 1. 
Step 2 learns  $\bf E$,
e.g., using HySime, step 3 and step 4 compute and denoise the eigen-images, respectively, and step 5 reconstructs the 
estimate of the original data.

%% Algorithm 1
\begin{algorithm}[h]  
\caption{FastHyDe  for i.i.d. noise}  
\label{code:alg1} 
\begin{algorithmic}[1]  
\State Input Noisy HSI data $\mathbf{Y}$ and the noise variance $\sigma^2$.
\State Learn the subspace $\mathbf{E}$ from observations $\bf Y$ and compute the noisy eigen-images $ \mathbf{E}^T\mathbf{Y}$. 
\State Compute ${\widehat {\bf Z}}$ by denoising the eigen-images  one by one,
using a state-of-the-art denoising algorithm.
\State Compute $\widehat{\mathbf{X}} = \mathbf{E \widehat{Z}}$, an estimate of the clean HSI.
 
\end{algorithmic}  
\end{algorithm}  

Note that the dimension of subspace ${\bf E}$ can be estimated by HySime or by any other subspace identification method. We provide evidence in the experiments that, as far as the subspace dimension is not underestimated, FastHyDe is extremely robust to errors in estimation of the subspace dimension.

\subsection{Additive Gaussian non-i.i.d. noise}
\label{sec:noniid}
Consider that in the observation model (\ref{obs}), $\mathbf{N}$ is zero-mean, additive,  Gaussian,
pixelwise independent with spectral covariance  $\mathbf{C}_\lambda = E[{\bf n}_i{\bf n}^T_i]$, where ${\bf n}_i$ is a generic column of $\bf N$. Notice that in the i.i.d. case 
$\mathbf{C}_\lambda = \sigma^2{\bf I}$, which is not the case in  the non-i.i.d. scenario.
We assume that ${\bf C}_{\lambda}$ is positive definite 
and therefore non-singular.
In order to reconvert the non-i.i.d. scenario into the i.i.d. one,  the observed data is whitened as
\begin{equation}
\label{eq:transfnoniid}
\mathbf{\widetilde{Y}} := \sqrt{\mathbf{C}_\lambda^{-1}} \mathbf{Y},
\end{equation}
where $\sqrt{\mathbf{C}_\lambda^{-1}}$ is a matrix denoting the square root of $\mathbf{C}_\lambda^{-1}$ and $\sqrt{\mathbf{C}_\lambda}$ denotes its inverse.
Then the observation model of $\mathbf{\widetilde{Y}}$ becomes
\begin{equation}
 \mathbf{\widetilde{Y}} = \sqrt{\mathbf{C}_\lambda^{-1}} \mathbf{X} + \sqrt{\mathbf{C}_\lambda^{-1}} \mathbf{N} = \mathbf{\widetilde{X}} + \mathbf{\widetilde{N}}.
\end{equation}
The noise covariance matrix of $\widetilde{\bf n}_i$, a generic column of
$\widetilde{\bf N}_i$,  is 
\begin{equation}
\label{eq:iid_converted}
\mathbf{\widetilde{C}}_\lambda = E[\widetilde{\bf n}_i\widetilde{\bf n}^T_i] = \mathbf{I}.
\end{equation}

Since the noise in \eqref{eq:iid_converted} is i.i.d., we may again formulate the denoising problem as 
\begin{equation}
\label{eq:noniid}
\mathbf{\widehat{\widetilde{Z}}}  = \arg\min_{\mathbf{\widetilde{Z}}} \frac{1}{2}\Big\| \mathbf{\widetilde{Z}-\widetilde{E}}^T{\bf \widetilde{Y}} \Big\|_F^2 +\lambda \phi (\mathbf{\widetilde{Z}}),
\end{equation}
where $\widetilde{\mathbf{E}}$ holds an orthonormal basis  learned from $\widetilde{\mathbf{Y}}$. The solution of (\ref{eq:noniid}) can be found following the same steps described in Algorithm \ref{code:alg1}. The clean data of $\widetilde{\mathbf{Y}}$ is estimated as
\begin{equation}
\label{eq:recov_x_td}
\widehat{\widetilde{\mathbf{X}}} =\widetilde{\mathbf{E}} \widehat{\widetilde{\mathbf{Z}}}, 
\end{equation}
followed by recovering the original clean data $\mathbf{X}$ from $\widetilde{\mathbf{X}}$
\begin{equation}
\label{eq:recov_x}
\widehat{\mathbf{X}} = \sqrt{\mathbf{C}_\lambda} \widehat{\widetilde{\mathbf{X}}}. 
\end{equation}

\section{Fast Inpainting}
\label{sec:fasthyin}

The clean image ${\bf X} \in {\mathbb R}^{n_b \times n}$ may be vectorized as ${\bf x}={\rm vec}({\bf X})\in \mathbb{R}^{t}$, $(t=n_b\times
n)$, where  $\rm vec$ stacks the columns of ${\bf X}$.  
In hyperspectral inpainting problems under the additive noise, the observation model may then be written as
\begin{equation}
\mathbf{y}=\mathbf{M}\mathbf{x}+ \mathbf{n}
\end{equation}
where  $\mathbf{y}\in \mathbb{R}^q$, with $q\leq t$, and $\mathbf{n} \in \mathbb{R}^q$ represents the observed incomplete data and noise, respectively. The matrix $\mathbf{M} \in \mathbb{R}^{q\times t}$ is a mask that selects a subset of the components of 
$\bf x$. It is thus a binary matrix corresponding to a subset of rows of an identity matrix, therefore having only one 1 per row. We assume that $\mathbf{M}$ is user-provided.

Below we start by considering Gaussian i.i.d. noise. Later, we consider Gaussian non-i.i.d and Poissonian noise.

\subsection{Additive Gaussian i.i.d. noise}
\label{subsec:inp_iid}
By still assuming that the spectral vectors belong to subspace spanned by the columns of $\bf E$, and thus 
, $\mathbf{X=EZ}$, we have
\begin{equation}
\mathbf{x}:= \rm vec ({\bf EZ}) = (\mathbf{I}\otimes \mathbf{E})\mathbf{z},
\end{equation}
where $\otimes$ represents Kronecker product and $\mathbf{z}:=\text{vec}(\mathbf{Z})$.
The inpainting problem is formulated as 
\begin{equation}
\label{eq:inp_z}
  \min_{\mathbf{z}} \frac{1}{2}\Vert \mathbf{M}(\mathbf{I} \otimes \mathbf{E})\mathbf{z}-\mathbf{y} \Vert^2 +\lambda \phi (\mathbf{z}).
\end{equation}

Instead of solving \eqref{eq:inp_z}, which involves a non-diagonal operator, thus, calling for iterative solvers,
we propose a suboptimal solution that  is very fast and effective. Let  ${\bf M}_i$ be the submatrix of $\bf M$ acting on  the pixel $i$, and
\begin{equation}
\label{eq:incomp_y_i}
{\bf y}_{o,i} = \mathbf{M}_i { {\bf x}}_i + { {\bf n}}_{o,i}
\in {\mathbb R}^{n_i},
\end{equation}
be the observed vector at the same pixel, where 
 $n_i \leq n_b$  denotes the number of observed components at that pixel, and   the vector ${ {\bf n}}_{o,i} \in {\mathbb R}^{n_i}$ denotes the corresponding noise components. Since ${\bf x}_i= {\bf Ez}_i$, we may write 
\begin{equation}
\label{eq:inp_ithpixel}
{\bf y}_{o,i} = \mathbf{M}_i {\bf Ez}_i + { {\bf n}}_{o,i}.
\end{equation}
The least squares estimator of coefficients of subspace representation ${ {\bf  z}}_i$ is given by
\begin{equation}
\label{eq:comp_z}
\widehat{{ {\bf z}}}_i = (\mathbf{E}^T \mathbf{M}_i^T \mathbf{M}_i \mathbf{E})^{-1} \mathbf{E}^T \mathbf{M}_i^T{ {\bf y}}_{o,i},
\end{equation}
where it is assumed that the matrix $\mathbf{E}^T \mathbf{M}_i^T \mathbf{M}_i \mathbf{E}$ is non-singular and thus that $n_i \geq k$. That is, the number of observed components is larger or equal than the dimension of the signal subspace.

After recovering the unobserved component at each pixel by computing  $\widehat{ {\bf y} }_i = \mathbf{E} \widehat{{ {\bf z}}}_i$, for $i=1, \dots, n$,  we solve the denoising problem 
\begin{equation}
 \label{eq:inp_y}
 \min_{\mathbf{z}} \frac{1}{2}\Vert  (\mathbf{I} \otimes \mathbf{E})\mathbf{z}-\widehat{\mathbf{y}} \Vert^2 +\lambda \phi (\mathbf{z}),
\end{equation}
where $\widehat{\mathbf{y}} := \text{vec}\,[\widehat{ {\bf y} }_1,\dots,\widehat{ {\bf y} }_n]$. Problem  \eqref{eq:inp_y}
is equivalent to
\begin{equation}
\label{eq:inp_Z}
   \min_{\mathbf{Z}} \frac{1}{2}\Vert \mathbf{EZ}-\widehat{\mathbf{Y}} \Vert_F^2 +\lambda \phi (\mathbf{Z}),
\end{equation}
Optimization \eqref{eq:inp_Z} is similar to  (\ref{equ:iid}) and, thus, we use FastHyDe to solve it.

At this point, we would like to remark that the pixels completely observed, that is $n_i = n_b$, are not affect by the computation of $\widehat{\bf z}_i$ given \eqref{eq:comp_z}. In fact, when $n_i = n_b$, ${\bf M}_i$ is a diagonal matrix and we have $\widehat{\bf z}_i = {\bf E}^T{\bf y}_i$, which is exactly the  vector used in the the  second equation in \eqref{equ:iid}. Therefore, the computation
\eqref{eq:comp_z} shall be applied just to the set of incomplete observed pixels.

The proposed approach for HSI inpainting, termed {\em Fast Hyperspectral inpainting} (FastHyIn), is summarized in Algorithm 2. 
As an   extension of FastHyDe,  FastHyIn turns into FastHyDe, when the inpainting mask is $\mathbf{M}=\mathbf{I}$.

%% Algorithm 1
\begin{algorithm}[h]  
\caption{FastHyIn for i.i.d. noise}  
\label{code:alg2} 
\begin{algorithmic}[1]  
\State  Input the noisy and incomplete HSI data $\mathbf{y}$ and the inpainting mask $\mathbf{M}$ 
\State Learn the subspace $\mathbf{E}$ from  completely observed pixels. 
\State Compute the eigen-images $\hat{ {\bf z}}_i$ according to (\ref{eq:comp_z}), only for incompletely observed pixels.
\State Compute ${\widehat {\bf Z}}$ by denoising the eigen-images  one by one,
using a state-of-the-art denoising algorithm.
\State Compute $\widehat{\mathbf{X}} = \mathbf{E \widehat{Z}}$, an estimate of the clean HSI.

\end{algorithmic}  
\end{algorithm}

\subsection{Additive Gaussian non-i.i.d. noise}
%\label{eq:inp_ithpixel}

We may treat the non-i.i.d noise as the maximum likelihood (ML) estimation of ${\bf z}_i$ in \eqref{eq:inp_ithpixel} under zero-mean Gaussian noise with correlation 
${\bf C}_{\lambda,i} = {\bf M}_i{\bf C}_{\lambda}{\bf M}_i^T$. The solution is 
\begin{equation}
\label{eq:comp_z_niid}
\widehat{{ {\bf z}}}_i = (\mathbf{E}^T \mathbf{M}_i^T
{\bf C}_{\lambda,i}^{-1} \mathbf{M}_i \mathbf{E})^{-1} \mathbf{E}^T \mathbf{M}_i^T
{\bf C}_{\lambda,i}^{-1}{ {\bf y}}_{o,i}.
\end{equation}
As in the i.i.d case, the coefficients \eqref{eq:comp_z_niid}
need to be computed just for the incompletely observed pixels.
Algorithm \ref{code:alg21} shows FastHyIn algorithm for non-i.i.d noise. 
%% Algorithm 21
\begin{algorithm}[h]  
	\caption{FastHyIn for non-i.i.d. noise}  
	\label{code:alg21} 
	\begin{algorithmic}[1]  
		\State  Input the noisy and incomplete HSI data $\mathbf{y}$ and the inpainting mask $\mathbf{M}$. 
		\State Learn the subspace $\mathbf{E}$ from  completely observed pixels.
		\State Compute $\widehat{ {\bf z}}_i$ according to (\ref{eq:comp_z_niid}), only for incompletely observed pixels, and replace
		${\bf y}_{o,i}$ with $\widehat{\bf y}_{i}= {\bf E}\widehat{ {\bf z}}_i$.
		\State Whiten the observations using \eqref{eq:transfnoniid}.
		\State Recompute $\widetilde{\bf E}$, a basis for 
		the subspace spanning the whitened data $\widetilde{\bf Y}$.
		
		\State Compute ${\widehat {\bf Z}}$ by denoising the eigen-images  one by one,
		using a state-of-the-art denoising algorithm.
		\State Compute $\widehat{\mathbf{X}} = \sqrt{\bf C}_\lambda\widetilde{\mathbf{E}} \widehat{\bf Z}$, an estimate of the clean HSI.

	\end{algorithmic}  
\end{algorithm}

Note that Algorithm \ref{code:alg21} is applied to general Gaussian non-i.i.d. cases where noise is pixelwise independent but, possibly, bandwise dependent, meaning covariance matrix $\mathbf{C}_\lambda$ can be non-diagonal.
Now let us consider a simpler non-i.i.d. case where noise is both  pixelwise and bandwise independent, meaning covariance matrix $\mathbf{C}_\lambda$ is diagonal. In this case, Algorithm \ref{code:alg21} can be simplified as follows: instead of implementing steps 1-4, we  whiten the incomplete observed pixel ${\bf y}_{o,i}$, that is, we compute
\begin{equation}
\tilde{{\bf y}}_{o,i} := \sqrt{\mathbf{C}_{\lambda,i}^{-1}}  {\bf y}_{o,i} ,
\label{eq:whit_inp}
\end{equation}
where ${\bf C}_{\lambda,i} = {\bf M}_i{\bf C}_{\lambda}{\bf M}_i^T$. The complete observed pixel can be also whiten using (\ref{eq:whit_inp}) with ${\bf M}_i={\bf I}$. After whitening data, we obtain $ \tilde{{\bf y}}  = [ \tilde{{\bf y}}_{o,1}^T, \tilde{{\bf y}}_{o,2}^T, \dots, \tilde{{\bf y}}_{o,n}^T]^T$, whose noise is i.i.d. We may again formulate the inpainting problem w.r.t $ \tilde{{\bf y}} $ with i.i.d. noise as in (\ref{eq:inp_z}).
 
\section{Evaluation with simulated data}
\label{sec:exp}
\subsection{Denoising experiments}
\label{sec:exp_de}

Two semi-synthetic observed hyperspectral datasets were simulated using additive Gaussian noise or Poissonian noise. The clean images were obtained form the  Pavia Centre subscene \footnote{\scriptsize Pavia scenes were provided by Prof. Paolo Gamba from the Telecommunications and Remote Sensing Laboratory, Pavia university (Italy) and can be downloaded from \url{http://www.ehu.eus/ccwintco/index.php?title=Hyperspectral_Remote_Sensing_Scenes}.} (of size $[200\times200]({\rm pixels})\times 103({\rm bands})$) 
and the Washington DC
Mall subscene \footnote{\scriptsize  This data set is available from the Purdue University Research Repository (\url{https://engineering.purdue.edu/~biehl/MultiSpec/hyperspectral.html})} (of size $[256\times 256]({\rm pixels}) \times 191({\rm bands})$).

\begin{figure*}[htbp]
\centering
\includegraphics[scale=0.5]{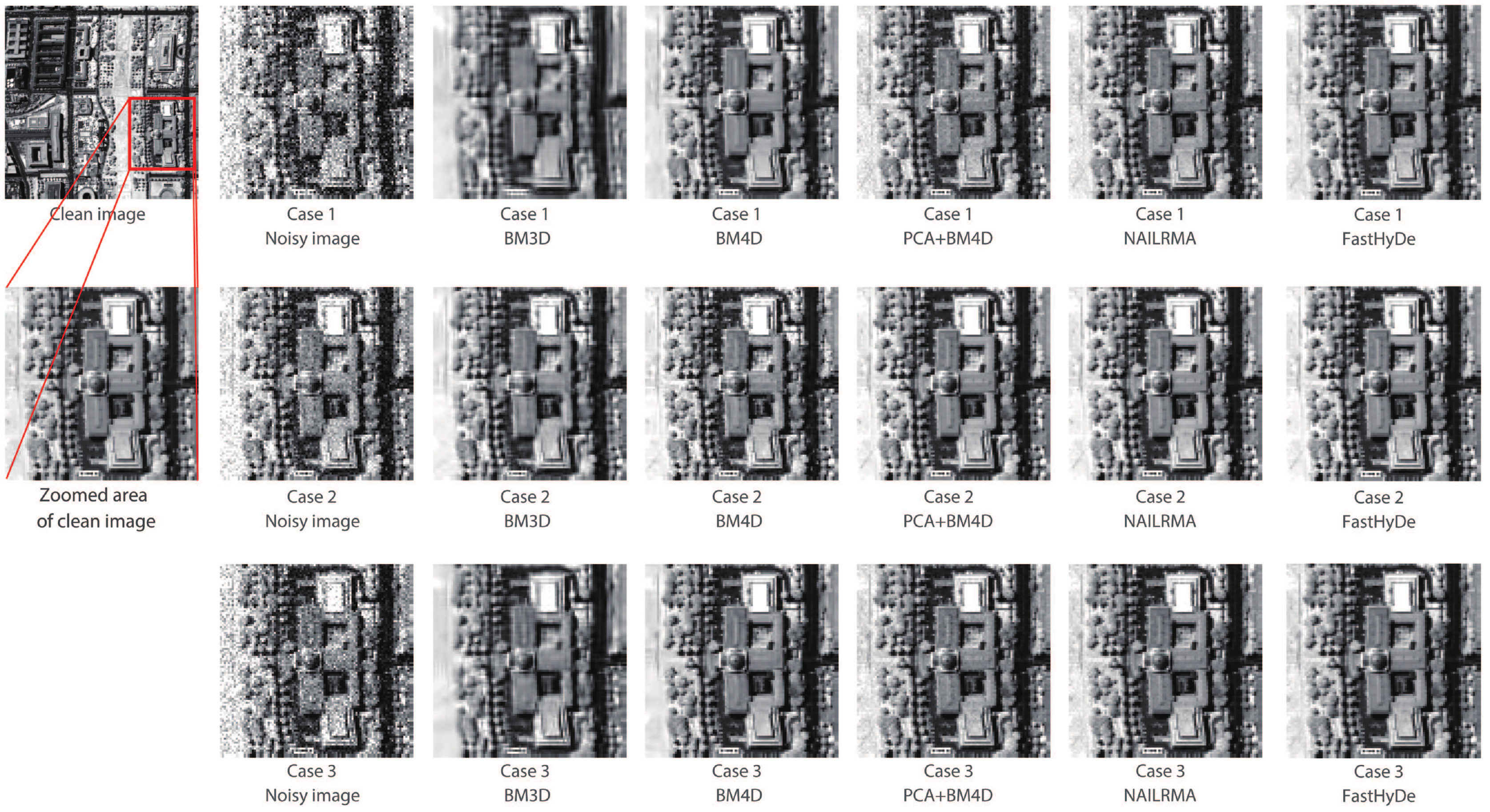}
\caption{Denoising result for  band 70 of DC Mall data with different kinds of noise.  Case 1: Gaussian i.i.d. noise. Case 2: Gaussian non-i.i.d. Case 3:  Poissonian noise}
\label{fig:dc_denoised}
\end{figure*}

\begin{figure*}[htbp]
\centering
\includegraphics[scale=0.5]{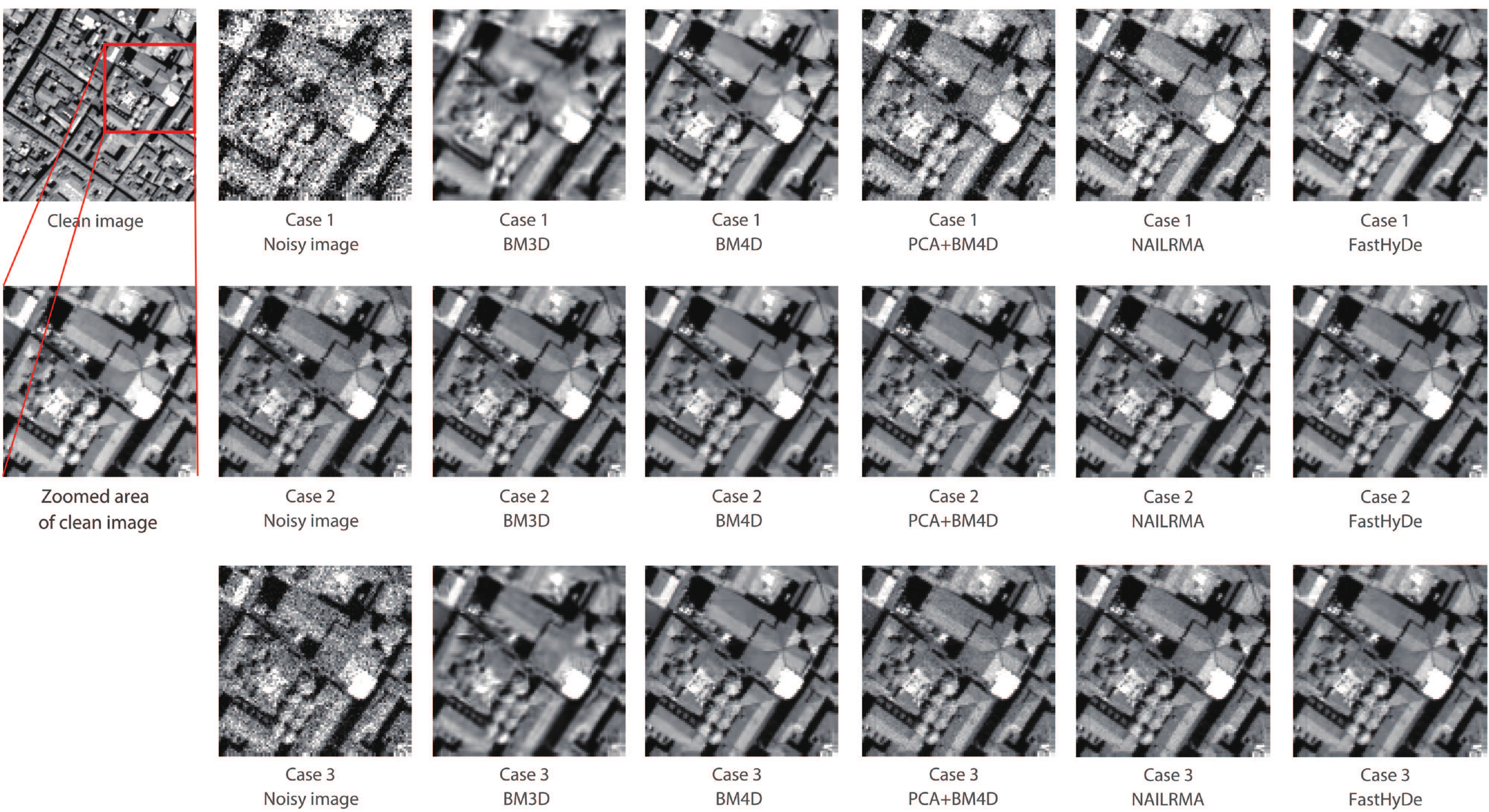}
\caption{Denoising result for  band 70 of Pavia Centre data with different kinds of noise.  Case 1: Gaussian i.i.d. noise. Case 2: Gaussian non-i.i.d. Case 3:  Poissonian noise.}
\label{fig:pavia_denoised}
\end{figure*}

%%%%%%%%%%%%%%%%%%%%%%%%%%%%%%%%%%%%%%%%%%%%%%%%%%%%%%%%
% vou aqui
%%%%%%%%%%%%%%%%%%%%%%%%%%%%%%%%%%%%%%%%%%%%%%%%%%%%%%%%
In order to simulate the clean images, 23 very low signal-to-noise bands, due to the water vapor absorption,    were removed.  The remaining spectral vectors, in the two data sets, were projected on a signal subspace learned via SVD. The dimensions of the signal subspace for the Washington DC Mall subscene and the Pavia Centre subscene were set to 8.  The projection
preserves the bulk of the signal energy and largely reduces the noise.  The projected HSIs of high quality are considered clean images in this  section. Before adding simulated noise, each band of clean HSIs are normalized to [0, 1] as in \cite{NAILRMA}. We remark that
the normalization is a linear and invertible operation that does not modify the signal-to-noise-ratio (SNR) of the corresponding band and that may be reverted after the denoising.
The band 70 of the clean images of the Washington DC
Mall subscene and of the Pavia Centre subscene   is shown in the top left hand side of Figs.
\ref{fig:dc_denoised} and \ref{fig:pavia_denoised}, respectively.

\vspace{0.2cm}
\noindent
Three kinds of additive noises are considered:
\begin{description}
	\item[Case 1] (Gaussian i.i.d. noise): ${\bf n}_i\sim {\cal N} ({\bf 0},{\bf \sigma}^2{\bf I}) $ with $\sigma\in\{0.02, 0.04, 0.06, 0.08, 0.10\}$.

	\item[Case 2] (Gaussian non-i.i.d. noise): ${\bf n}_i\sim {\cal N} ({\bf 0},{\bf D}^2) $ where $\bf D$ is a diagonal matrix with diagonal elements sampled from a Uniform distribution $U(0,1)$.
	
	\item[Case 3] (Poissonian noise): ${\bf Y} \sim{\cal P}(\alpha {\bf X})$, where  ${\cal P}(\bf A)$ stands for a matrix of size$({\bf A})$ of independent Poisson random variables  whose parameters are given by the corresponding element of ${\bf A}:=[a_{ij}]$. The parameter $\alpha$ is such that  
	SNR$\,:=\alpha(\sum_{i,j}a_{ij}^2)/(\sum_{ij}a_{ij})$  was set 15 dB.

\end{description}

The proposed FastHyDe \footnote{\textcolor{red}{Matlab code of FastHyDe is available in  \url{www.lx.it.pt/~bioucas/code/Demo_FastHyDe_FastHyIn.rar}}}
 is compared with  BM3D \cite{BM3D}, applied band by band,  BM4D \cite{BM4D}, `PCA+BM4D' \cite{pca+vbm4d}, and NAILRMA \cite{NAILRMA}. Since all the compared methods assume i.i.d. noise, the observed data in Case 2 and Case 3 are transformed in order to  to have additive i.i.d. noise before the denoisers are applied. In case 2, we applied the transformation (\ref{eq:transfnoniid}). In case 3, we applied the Anscombe transform $\mathbf{\widetilde{Y}} :=2\sqrt{\mathbf{Y}+\frac{3}{8}}$, which converts 
Poissonian noise into approximately additive  noise \cite{AnscT}.

We now discuss the parameter setting of the various algorithms. The parameters of FastHyDe and FastHyIn  are the noise covariance matrix ${\bf C}_\lambda$ and  the signal subspace, which are both computed by HySime \cite{Hysime}. BM3D and BM4D require the value of standard deviation of the noise as input parameter, which is estimated by HySime \cite{Hysime} as well. The remaining  BM3D and BM4D parameters, namely the patch size $N1$, the sliding step $Nstep$, and the size of the search neighborhood $Ns$, are set to the  default values stated in \cite{BM3D,BM4D}:  $N1$=4, $Nstep$=3 $Ns$=11, for BM3D and, $N1$=8, $Nstep$=3, and $Ns$=39, for BM4D.
The subspace dimension of two datasets input to FastHyDe and FastHyIn was set to 10, instead of 8, the true value, to provide evidence of the robustness of FastHyDe and FastHyIn with respect to subspace overestimation.  The subspace dimension input to `PCA+BM4D' was set to 8. For NAILRMA, the block size was set to 20 and step size to 8. These values were hand tuned to get optimal performance.

\begin{figure*}[htbp]
\centering
\includegraphics[scale=1.15]{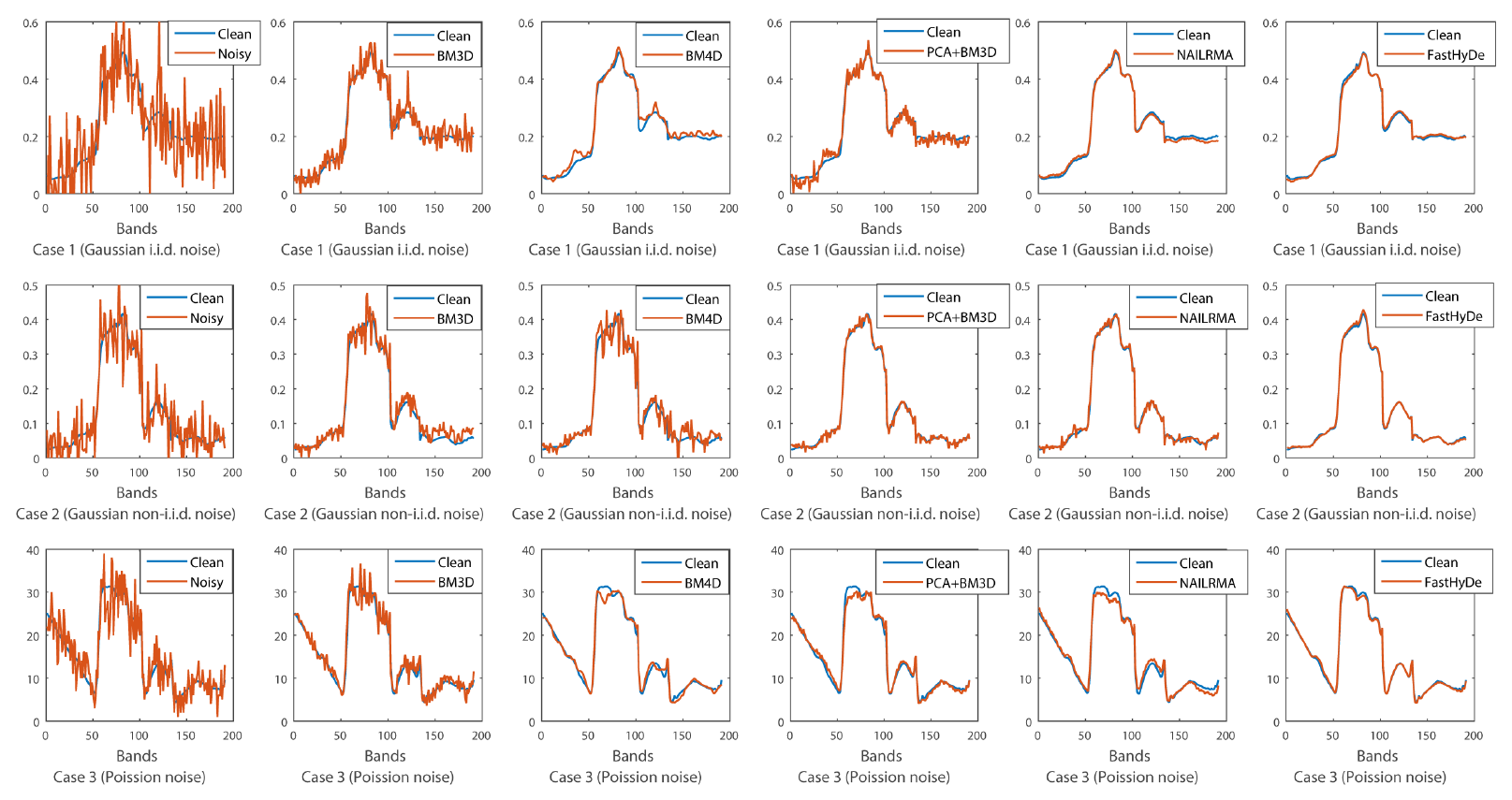}
\caption{Denoised spectral signatures of Washington DC Mall data with different kinds of noise.
}
\label{fig:specLine_dc}
\end{figure*}

\begin{figure*}[htbp]
\centering
\includegraphics[scale=1.15]{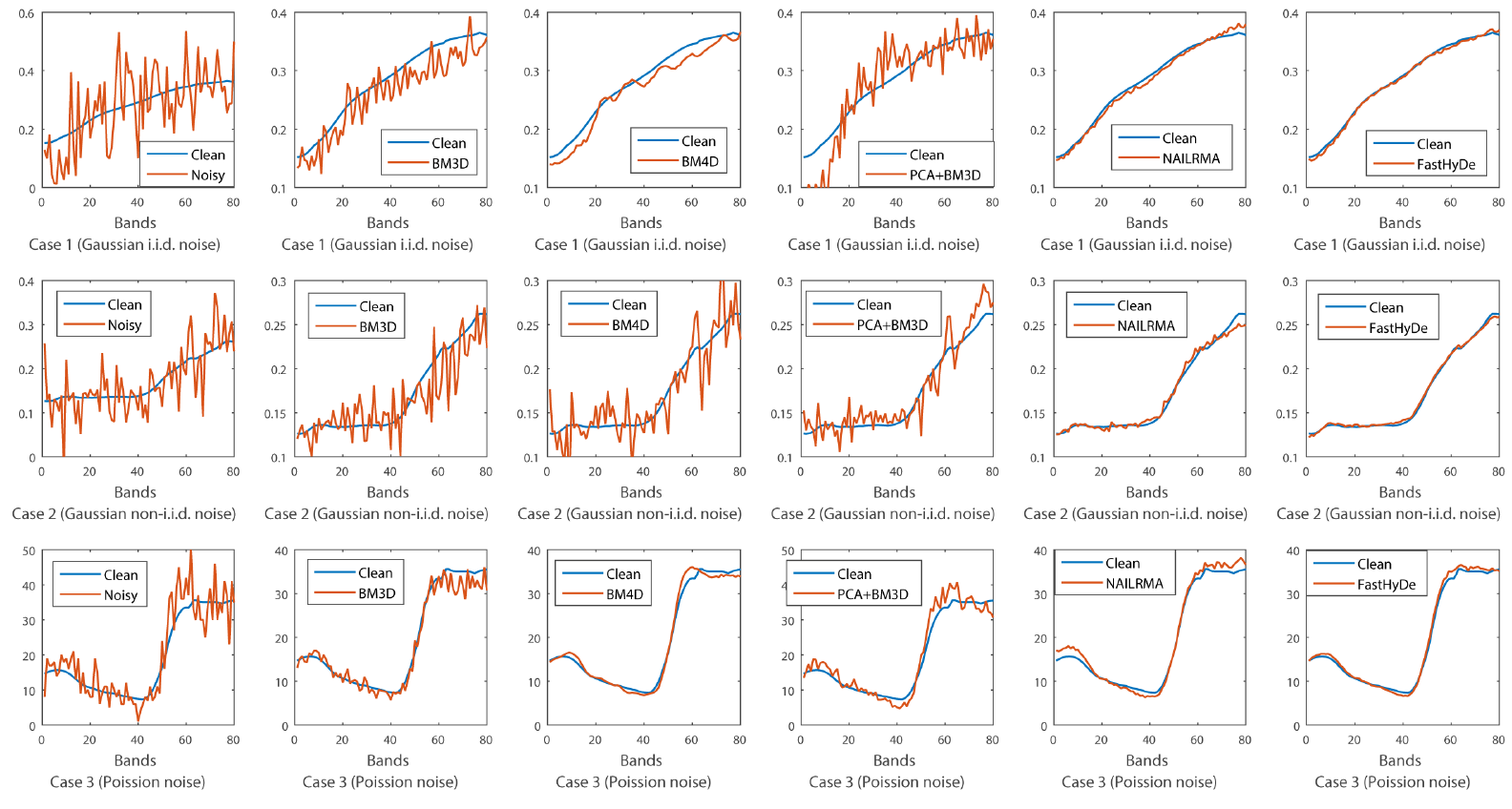}
\caption{  Denoised spectral signatures of Pavia Centre data with different kinds of noise. }
\label{fig:specLine_pavia}
\end{figure*}

For quantitative assessment, the peak signal-to-noise (PSNR) index and the structural similarity (SSIM) index of each band are calculated. The corresponding mean PSNRs (MPSNR) and mean SSIMs (MSSIM) in Washington DC Mall data and in Pavia Centre are reported in \cref{tab:wash,tab:pav}, respectively.  Dealing with different kinds of noises, FastHyDe yields uniformly the  best performance with gains increasing as the noise  increases, as it may be concluded from those results. High quality spectral signature is of critical importance to material identification. 
The quality of reconstructed spectra from different denoising methods may also be inferred  from  \cref{fig:dc_denoised,fig:pavia_denoised,fig:specLine_dc,fig:specLine_pavia}.

BM3D suits very well the  eigen-images, which are self-similar. In addition, the fact that denoising is applied only to the  eigen-images, which are much less than the number of bands,  significantly reduces the FastHyDe complexity (see \cref{tab:washtime,tab:paviatime}),  relative to the competitors. The algorithms were implemented using MATLAB R2010 on a desktop PC equipped with eight Intel Core i7-4970 CPU (at 3.60 GHz) and 16 GB of RAM memory.

\begin{figure}[htbp]
\centering
\includegraphics[scale=0.45]{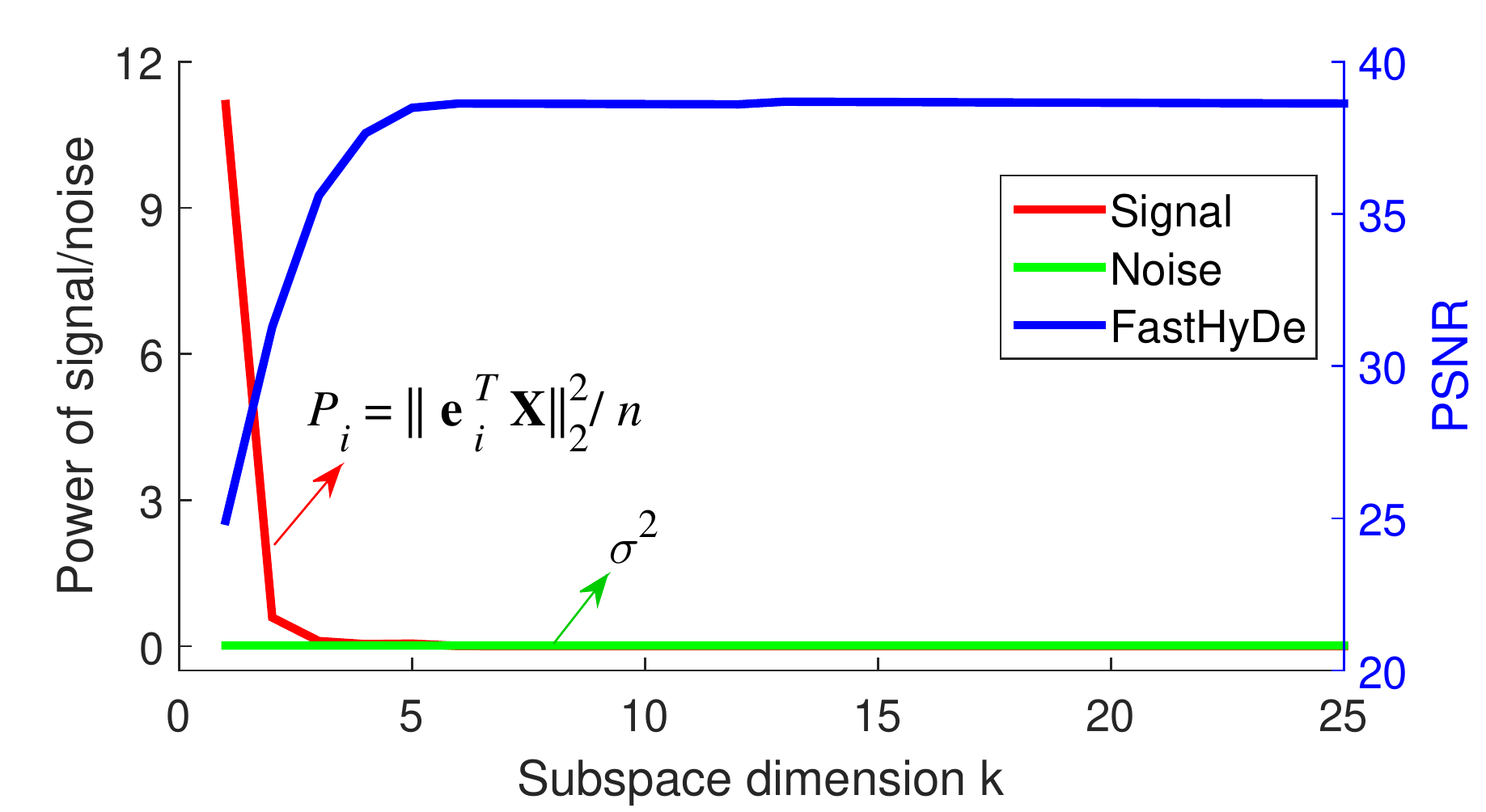}
\caption{{Power of signal and of noise and  FastHyDe PSNR as a function of the  dimension of the subspace estimation (the true subspace dimension is 8).}}
\label{fig:ImpaceOfk}
\end{figure}

FastHyDe robustness to subspace overestimation is illustrated   in Fig \ref{fig:ImpaceOfk}. 
Take as example the noisy image  of Washington DC Mall data generated in case 1 with $\sigma = 0.10$ (Gaussian i.i.d. noise).
The curve in red represents the energy of the original data $\bf X$ along the eigenvalue directions, that is, the values of $\| {\bf e}_i^T{\bf X}\|_2^2/n$, ordered by non-increasing magnitude. The straight line in green  represents the noise variance. The curve in blue represents the PSNR yielded by FastHyDe for different values of the subspace dimension. It is clear that the PSNR is  practically constant provided that the subspace dimension in not underestimated.  That is to say that FastHyDe is extremely robust to the subspace dimension overestimation.

\begin{figure*}[htbp]
\centering
\includegraphics[scale=0.96]{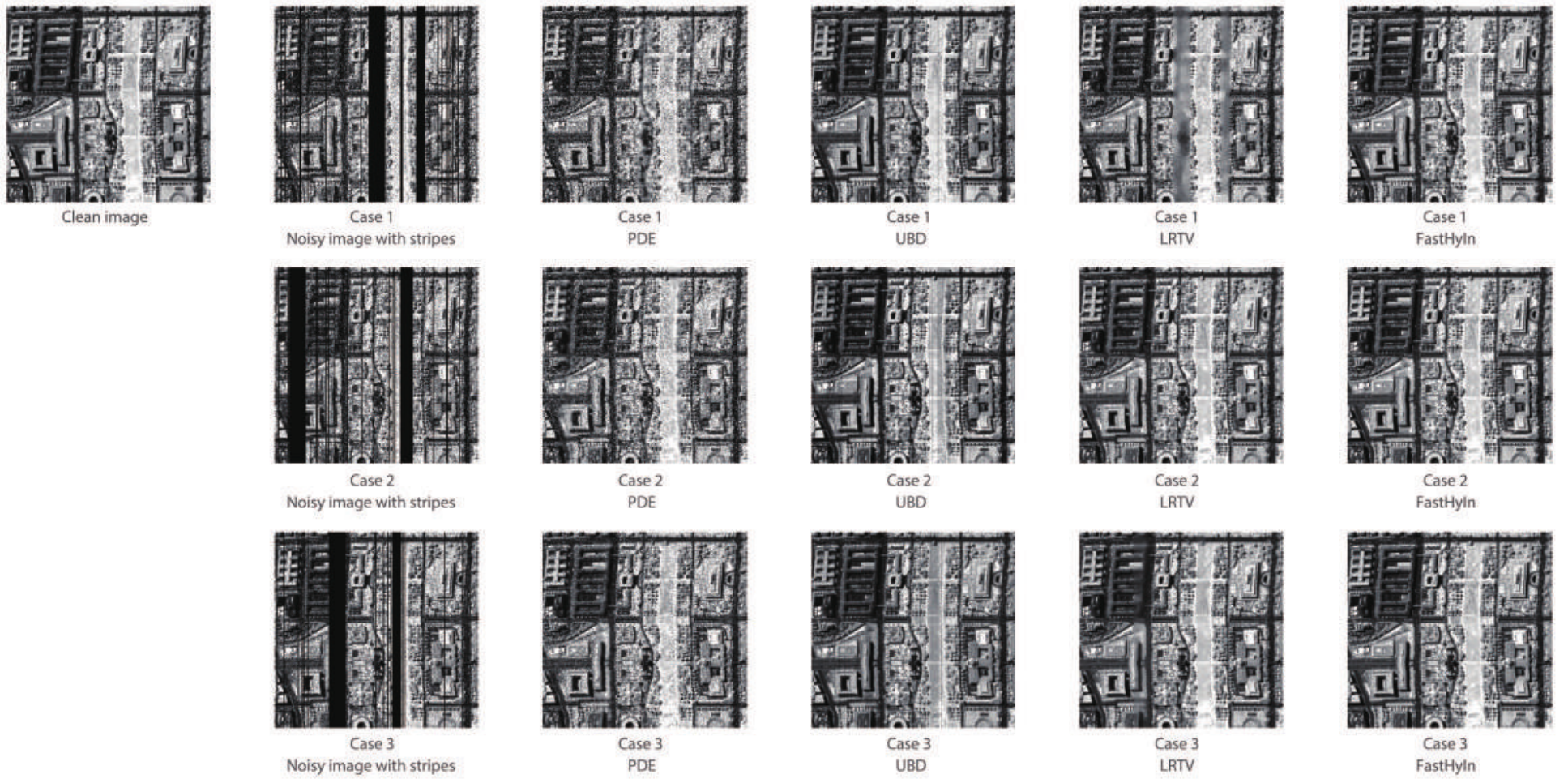}
\caption{Destriping band 60 of  Washington DC Mall with different kinds of noise.  Case 1: Gaussian i.i.d. noise. Case 2: Gaussian non-i.i.d. Case 3:  Poissonian noise.}
\label{fig:dc_inpainted}
\end{figure*}

\begin{figure*}[htbp]
\centering
\includegraphics[scale=0.96]{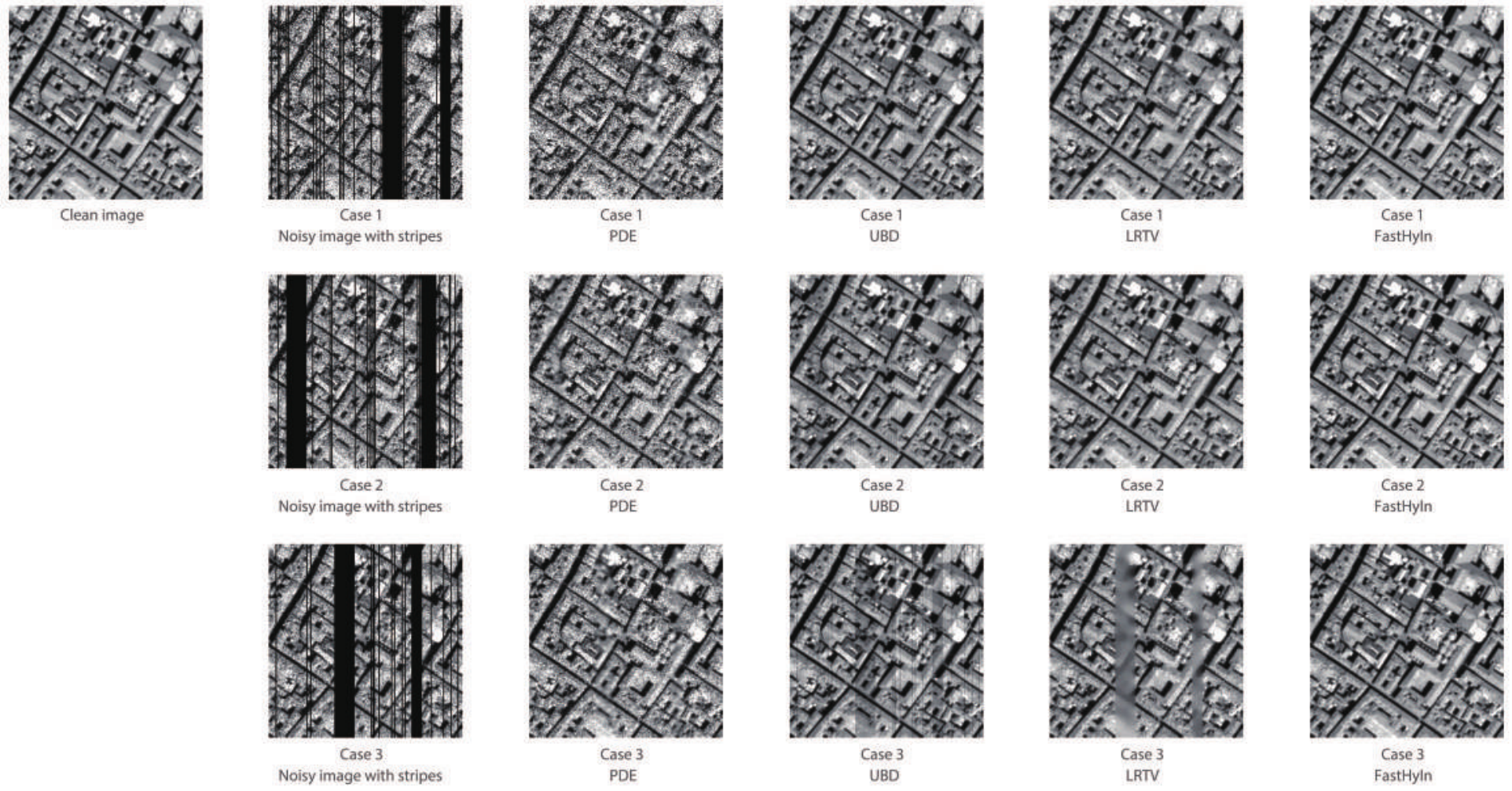}
\caption{Destriping band 60 of  Pavia Centre with different kinds of noise.  Case 1: Gaussian i.i.d. noise. Case 2: Gaussian non-i.i.d. Case 3:  Poissonian noise.}
\label{fig:pavia_inpainted}
\end{figure*}

% Table generated by Excel2LaTeX from sheet 'EqualNoise'
\begin{table*}[htbp]
%\small
  \centering
  \caption{Quantitative assessment of different denoising algorithms applied to Washington DC Mall.}
    \begin{tabular}{ccccccccc}
    \toprule
      & $\sigma$ & Index & Noisy Image & BM3D & BM4D & PCA+BM4D & NAILRMA & FastHyDe \\
    \midrule
    \multirow{10}[2]{*}{Case 1} & \multirow{2}[1]{*}{0.02} & MPSNR(dB) & 33.98 & 36.25 & 44.39 & 46.90 & 48.37 & \textbf{49.70} \\
      &   & MSSIM & 0.9370 & 0.9663 & 0.9945 & 0.9960 & 0.9974 & \textbf{0.9981} \\
      & \multirow{2}[0]{*}{0.04} & MPSNR(dB) & 27.96 & 31.97 & 39.70 & 40.85 & 43.16 & \textbf{44.78} \\
      &   & MSSIM & 0.8098 & 0.9166 & 0.9840 & 0.9845 & 0.9920 & \textbf{0.9945} \\
     & \multirow{2}[0]{*}{0.06} & MPSNR(dB) & 24.44 & 29.77 & 36.22 & 37.34 & 40.24 & \textbf{42.00} \\
      &   & MSSIM & 0.6827 & 0.8677 & 0.9651 & 0.9666 & 0.9849 & \textbf{0.9902} \\
      & \multirow{2}[0]{*}{0.08} & MPSNR(dB) & 21.94 & 28.33 & 35.03 & 34.79 & 38.22 & \textbf{40.11} \\
      &   & MSSIM & 0.5728 & 0.8253 & 0.9536 & 0.9435 & 0.9769 & \textbf{0.9854} \\
      & \multirow{2}[1]{*}{0.10} & MPSNR(dB) & 20.00 & 27.29 & 33.48 & 32.78 & 36.67 & \textbf{38.57} \\
      &   & MSSIM & 0.4821 & 0.7848 & 0.9346 & 0.9169 & 0.9686 & \textbf{0.9801} \\
      \midrule
    \multirow{2}[1]{*}{Case 2} &   & MPSNR(dB) & 28.63 & 33.00 & 34.20 & 44.89 & 47.43 & \textbf{51.75} \\
      &   & MSSIM & 0.7463 & 0.8942 & 0.9295 & 0.9893 & 0.9971 & \textbf{0.9987} \\
      \midrule
    \multirow{2}[1]{*}{Case 3} &   & MPSNR(dB) & 26.98 & 31.29 & 38.82 & 39.56 & 41.75 & \textbf{43.21} \\
      &   & MSSIM & 0.7993 & 0.9118 & 0.9814 & 0.9804 & 0.9888 & \textbf{0.9919} \\
    \bottomrule
    \end{tabular}%
  \label{tab:wash}%
\end{table*}%

 % Table generated by Excel2LaTeX from sheet 'pavia'
\begin{table*}[htbp]
  \centering
  \caption{Quantitative assessment of different denoising algorithms applied to Pavia Centre data.}
    \begin{tabular}{ccccccccc}
    \toprule
      & $\sigma$ & Index & Noisy Image & BM3D & BM4D & PCA+BM4D & NAILRMA & FastHyDe \\
    \midrule
    \multirow{10}[2]{*}{Case 1} & \multirow{2}[1]{*}{0.02} & MPSNR(dB) & 33.98 & 36.67 & 44.58 & 43.09 & 46.66 & \textbf{47.09} \\
      &   & MSSIM & 0.9328 & 0.9663 & 0.9939 & 0.9903 & 0.9959 & \textbf{0.9965} \\
      & \multirow{2}[0]{*}{0.04} & MPSNR(dB) & 27.96 & 32.68 & 39.98 & 37.03 & 41.71 & \textbf{42.58} \\
      &   & MSSIM & 0.7950 & 0.9212 & 0.9832 & 0.9632 & 0.9874 & \textbf{0.9908} \\
      & \multirow{2}[0]{*}{0.06} & MPSNR(dB) & 24.44 & 30.56 & 37.39 & 33.45 & 38.87 & \textbf{40.03} \\
      &   & MSSIM & 0.6550 & 0.8770 & 0.9701 & 0.9227 & 0.9770 & \textbf{0.9840} \\
      & \multirow{2}[0]{*}{0.08} & MPSNR(dB) & 21.94 & 29.11 & 35.55 & 30.93 & 36.80 & \textbf{38.32} \\
      &   & MSSIM & 0.5362 & 0.8350 & 0.9556 & 0.8753 & 0.9646 & \textbf{0.9769} \\
      & \multirow{2}[1]{*}{0.10} & MPSNR(dB) & 20.00 & 28.06 & 34.11 & 29.00 & 35.31 & \textbf{36.99} \\
      &   & MSSIM & 0.4417 & 0.7966 & 0.9393 & 0.8251 & 0.9541 & \textbf{0.9700} \\
       \midrule
    \multirow{2}[2]{*}{Case 2} &   & MPSNR(dB) & 29.77 & 34.29 & 35.05 & 39.70 & 46.94 & \textbf{49.26} \\
      &   & MSSIM & 0.7445 & 0.9054 & 0.9225 & 0.9640 & 0.9970 & \textbf{0.9979} \\
       \midrule
    \multirow{2}[2]{*}{Case 3} &   & MPSNR(dB) & 26.97 & 32.18 & 39.46 & 35.95 & 40.53 & \textbf{41.59} \\
      &   & MSSIM & 0.7595 & 0.9109 & 0.9817 & 0.9522 & 0.9848 & \textbf{0.9890} \\
    \bottomrule
    \end{tabular}%
  \label{tab:pav}%
\end{table*}%

% Table generated by Excel2LaTeX from sheet 'EqualNoise'
\begin{table}[htbp]
 \footnotesize
  \centering
  \caption{Computational time (seconds) of different denoising algorithms applied to Washington DC Mall data.}
    \begin{tabular}{rcrrcrr}
    \toprule
    \multicolumn{1}{c}{} & \multirow{2}[2]{*}{$\sigma$} & \multicolumn{1}{r}{\multirow{2}[2]{*}{BM3D}} & \multicolumn{1}{r}{\multirow{2}[2]{*}{BM4D}} & PCA+ & \multicolumn{1}{r}{\multirow{2}[2]{*}{NAILRMA}} & \multicolumn{1}{r}{\multirow{2}[2]{*}{FastHyDe}} \\
   
      &   &   &   & BM4D &   &  \\
       \midrule
   
    \multicolumn{1}{c}{\multirow{5}[2]{*}{Case 1}} & 0.02 & \multicolumn{1}{c}{196} & \multicolumn{1}{c}{9679} & 9853 & \multicolumn{1}{c}{347} & \multicolumn{1}{c}{\textbf{15}} \\
    \multicolumn{1}{c}{} & 0.04 & \multicolumn{1}{c}{208} & \multicolumn{1}{c}{9883} & 9588 & \multicolumn{1}{c}{236} & \multicolumn{1}{c}{\textbf{14}} \\
    \multicolumn{1}{c}{} & 0.06 & \multicolumn{1}{c}{223} & \multicolumn{1}{c}{9698} & 8485 & \multicolumn{1}{c}{171} & \multicolumn{1}{c}{\textbf{15}} \\
    \multicolumn{1}{c}{} & 0.08 & \multicolumn{1}{c}{205} & \multicolumn{1}{c}{8771} & 8405 & \multicolumn{1}{c}{149} & \multicolumn{1}{c}{\textbf{14}} \\
    \multicolumn{1}{c}{} & 0.10 & \multicolumn{1}{c}{206} & \multicolumn{1}{c}{8865} & 8660 & \multicolumn{1}{c}{133} & \multicolumn{1}{c}{\textbf{13}} \\
     \midrule
    \multicolumn{1}{c}{Case 2} &   & \multicolumn{1}{c}{198} & \multicolumn{1}{c}{8867} & 8481 & \multicolumn{1}{c}{27} & \multicolumn{1}{c}{\textbf{16}} \\
     \midrule
    \multicolumn{1}{c}{Case 3} &   & \multicolumn{1}{c}{183} & \multicolumn{1}{c}{8899} & 8537 & \multicolumn{1}{c}{39} & \multicolumn{1}{c}{\textbf{28}} \\
    
    \bottomrule
    \end{tabular}%
\label{tab:washtime}%
\end{table}%

% Table generated by Excel2LaTeX from sheet 'pavia'
\begin{table}[htbp]
 \footnotesize
  \centering
  \caption{Computational time (seconds) of different denoising algorithms applied to Pavia Centre data.}
    \begin{tabular}{ccccccc}
    \toprule
      & \multirow{2}[2]{*}{$\sigma$} & \multirow{2}[2]{*}{BM3D} & \multirow{2}[2]{*}{BM4D} & PCA+ & \multirow{2}[2]{*}{NAILRMA} & \multirow{2}[2]{*}{FastHyDe} \\
   
      &   &   &   & BM4D &   &  \\
       \midrule
   
    \multirow{5}[2]{*}{Case 1} & 0.02 & 56 & 2118 & 1824 & 112 & \textbf{11} \\
      & 0.04 & 56 & 2045 & 1773 & 72 & \textbf{9} \\
      & 0.06 & 57 & 1995 & 1764 & 56 & \textbf{9} \\
      & 0.08 & 59 & 2000 & 1753 & 51 & \textbf{9} \\
      & 0.10 & 60 & 2004 & 1727 & 46 & \textbf{10} \\
        \midrule
    Case 2 &   & 48 & 1792 & 1607 & 9 & \textbf{7} \\
      \midrule
    Case 3 &   & 48 & 2189 & 1626 & 10 & \textbf{9} \\

    \bottomrule
    \end{tabular}%
 \label{tab:paviatime}%
\end{table}%

\subsection{Inpainting experiments}
\label{sec:exp_in}
This section evaluates the performance of FastHyIn  \footnote{ Matlab code of FastHyIn is available in  \url{www.lx.it.pt/~bioucas/code/Demo_FastHyDe_FastHyIn.rar} }
by comparing it with three inpainting methods: the partial differential equations discretization (PDE)  adapted to 3D data  \cite{pdeURL,PDE},  the unmixing based denoising (UBD) \cite{UBD},  and the total variation (TV)-regularized low-rank matrix factorization (LRTV) \cite{LR_zhang2}.
PDE method fills in selected regions with information surrounding them. UBD reconstructs damaged pixel using spectral unmixing results.   LRTV models stripes as sparse noise, exploits spectral low-rank property via nuclear norm, and promotes spatial piecewise smoothness via TV regularization. 
In our experiments, spectral unmixing in UBD method is implemented as follows: 
The number of endmembers, their spectral signatures, and their abundances at each pixel are estimated by 
HySime \cite{Hysime}, vertex component analysis \cite{VCA} and  non-negative constrained least squares, respectively.

The noisy images generated in case 1 with $\sigma = 0.10$ (Gaussian i.i.d. noise), case 2 (Gaussian non-i.i.d) and case 3 (Poissonian noise) (see more details in Section~\ref{sec:exp_de}) are used again to simulate corrupted images with stripes. Stripes are simulated for four bands (60th to 63rd bands) as shown in second columns in \cref{fig:dc_inpainted,fig:pavia_inpainted}. 
 The location of missing values is known beforehand and is an input variable in PDE, LRTV, and FastHyIn. 
 
Destriping results are shown in \cref{fig:dc_inpainted,fig:pavia_inpainted}.  Visually, for Washington DC Mall data, almost all the methods are able to inpaint the dead lines, except LRTV in Case 1.  For Pavia Centre data and in case 3,  UBD and LRTV methods do not remove the stripes completely, as shown in Fig.\ref{fig:pavia_inpainted}.

We remark that it is reasonable that output images of PDE are still noisy, because this method functions only as an inpainter and not as a denoiser.
In contrast with this scenario, UBD, LRTV and FastHyIn remove noise largely while inpainting. 
 Note that UBD reconstructs a clean image with estimated endmembers and abundances, meaning that its performance strongly depends on the results of spectral unmixing. Spectral unmixing is still a challenging problem in the realm of hyperspectral image processing \cite{overview,keshava2002spectral,plaza2009recent}.
For applying LRTV, the main challenge is to choose suitable regularization  parameters. We have hand tuned  them to get optimal performance in these experiments. We set rank $r$ = 8 (true value), the TV regularization parameter $\tau = 0.015$ for all images,  the sparsity regularization parameter $\lambda = 70/{\sqrt{n}}$ for Pavia Centre data, and $\lambda = 100/{\sqrt{n}}$ for Washington DC Mall data. FastHyIn is user-friendly. Similarly to FastHyDe, it involves only the  
noise covariance matrix ${\bf C}_\lambda$  and the signal subspace, which are both compute by HySime \cite{Hysime}. To provide evidence of FastHyIn robustness to overestimation of the subspace dimension, we set the subspace dimension $k=10$ instead of $k=8$, the true value.  Regarding removal of noise, we can see from  \cref{tab:inp_dc,tab:inp_pavia} that FastHyIn yields uniformly the best results in terms of MPSNR and MSSIM.

% Table generated by Excel2LaTeX from sheet 'Sheet1'
\begin{table*}[htbp]
  \centering
  \caption{Quantitative assessment of different inpainting algorithms applied to Washington DC Mall.}
   \begin{tabular}{ccccccc}
    \toprule
      & Index & Noisy Image & PDE & UBD & LRTV & FastHyIn \\
    \midrule
    \multirow{3}[0]{*}{Case 1} & MPSNR & 19.90 & 20.01 & 34.52 & 35.53 & \textbf{38.58} \\
      & MSSIM & 0.4794 & 0.4831 & 0.9575 & 0.9535 & \textbf{0.9802} \\
      & Time &  -  & 26 & 23 & 210 & \textbf{12} \\
    \multirow{3}[0]{*}{Case 2} & MPSNR & 28.43 & 28.63 & 35.46 & 41.88 & \textbf{51.63} \\
      & MSSIM & 0.7414 & 0.7466 & 0.9648 & 0.9853 & \textbf{0.9987} \\
      & Time &  -  & 35 & 23 & 210 & \textbf{13} \\
    \multirow{3}[1]{*}{Case 3} & MPSNR & 26.80 & 26.99 & 37.18 & 40.78 & \textbf{43.21} \\
      & MSSIM & 0.7952 & 0.7996 & 0.9780 & 0.9831 & \textbf{0.9919} \\
      & Time &  -  & 26 & 26 & 179 & \textbf{24} \\
    \bottomrule
    \end{tabular}%
  \label{tab:inp_dc}%
\end{table*}%

% Table generated by Excel2LaTeX from sheet 'pavia'
\begin{table*}[htbp]
  \centering
  \caption{Quantitative assessment of different inpainting algorithms applied to Pavia Centre data.}
       \begin{tabular}{ccccccc}
    \toprule
      & Index & Noisy Image & PDE & UBD & LRTV & FastHyIn \\
    \midrule
    \multirow{3}[1]{*}{Case 1} & MPSNR & 19.79 & 20.04 & 33.68 & 33.38 & \textbf{36.95} \\
      & MSSIM & 0.4335 & 0.4441 & 0.9335 & 0.9103 & \textbf{0.9698} \\
      & Time &  -  & \textbf{1} & 12 & 49 & 6 \\
    \multirow{3}[0]{*}{Case 2} & MPSNR & 28.93 & 29.40 & 35.73 & 39.62 & \textbf{48.60} \\
      & MSSIM & 0.7308 & 0.7454 & 0.9629 & 0.9565 & \textbf{0.9975} \\
      & Time &  -  & \textbf{1} & 11 & 49 & 6 \\
    \multirow{3}[1]{*}{Case 3} & MPSNR & 26.51 & 27.00 & 35.90 & 38.58 & \textbf{41.54} \\
      & MSSIM & 0.7453 & 0.7604 & 0.9647 & 0.9649 & \textbf{0.9889} \\
      & Time &  -  & \textbf{1} & 13 & 49 & 7 \\
    \bottomrule
    \end{tabular}%
  \label{tab:inp_pavia}%
\end{table*}%

\section{Evaluation with Real data}
\label{sec:exp_real}

\subsection{Denoising experiments}

 \begin{figure}[htbp]
\centering
\includegraphics[scale=3]{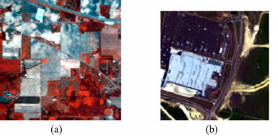}
\caption{False color composite of Indian Pine data (a) and Urban subscene (b).}
\label{fig:RealImg}
\end{figure}

\begin{figure*}[htbp]
\centering
\includegraphics[scale=0.8]{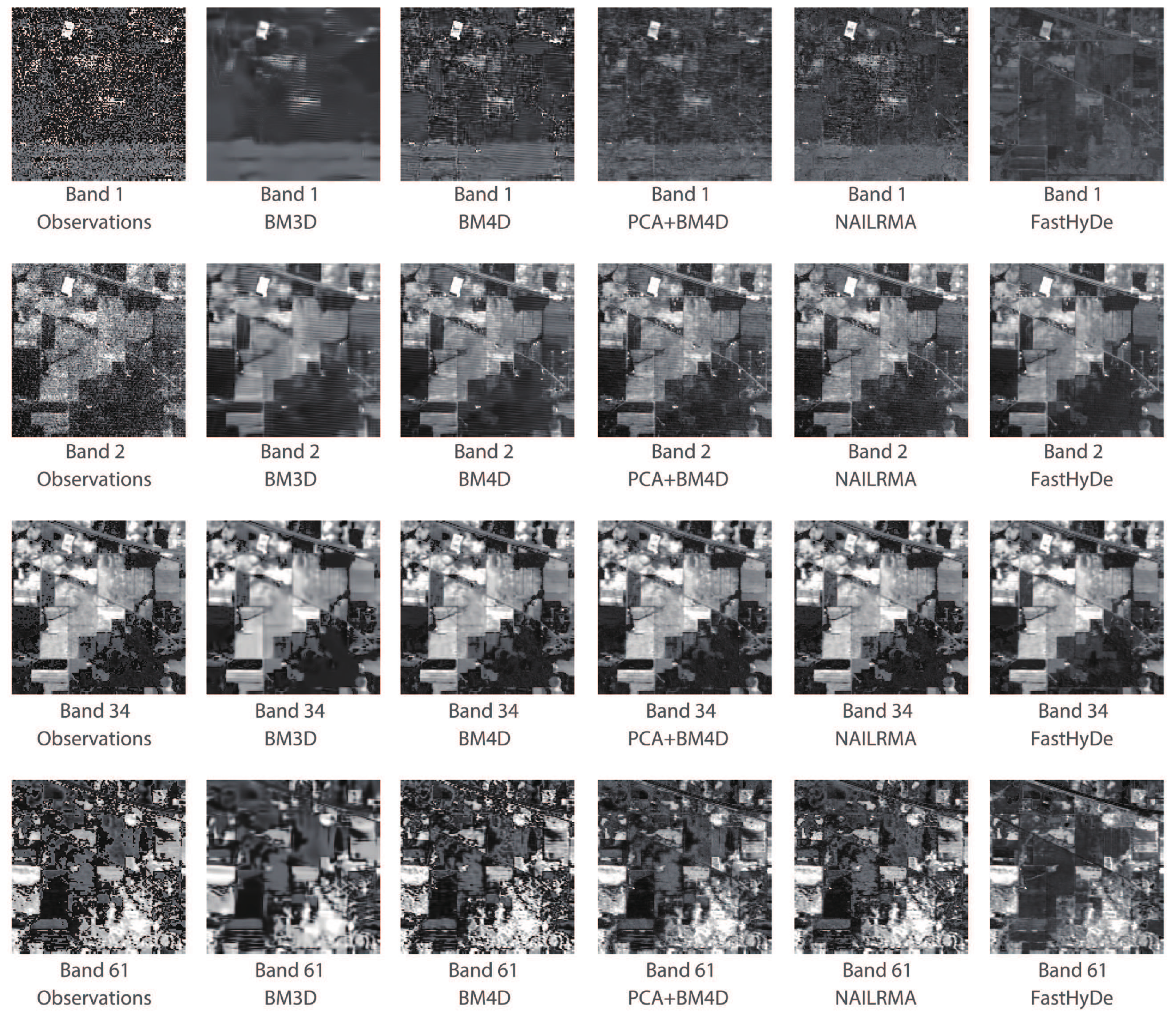}
\caption{Denoising results of BM3D (75 s), BM4D (2934 s), PCA+BM4D (2761 s), NAILRMA (11 s) and FastHyDe (10 s) in Indian Pine data.}
\label{fig:indPin_rgb}
\end{figure*}

In this section we apply FastHyde to the AVIRIS (airborne visible/infrared imaging spectrometer) Indian Pines scene (Fig. \ref{fig:RealImg}-(a)) \cite{PURR1947}. This image, recorded over North-western Indiana in June 1992, has $145 \times 145$ pixels, with a spatial resolution of 20 meters per pixel and 220 spectral channels. The image  displays  strong  noise in a number of bands.  The noise  was assumed to be non-i.i.d. and estimated 
with HySime.  The first column in Fig. \ref{fig:indPin_rgb} shows four of those noise bands (1, 2, 34, and 61). 

Indian Pine image was denoised by BM3D, BM4D, `PCA+BM4D', NAILRMA and FastHyDe. 
 The parameters used in BM3D and BM4D were set as in the simulated data. The subspace dimension input to `PCA+BM4D' was 18, which is estimated by HySime \cite{Hysime}. Considering FastHyDe is robust to subspace dimension overestimation, subspace dimension input to FastHyDe was set to 25. 
The results are exhibited in Fig. \ref{fig:indPin_rgb} and the corresponding computational times are reported in the figure caption. Qualitatively, FastHyDe  yields the best result, in the shortest time.

\subsection{Inpainting experiments}
 \begin{figure*}[htbp]
\centering
\includegraphics[scale=0.8]{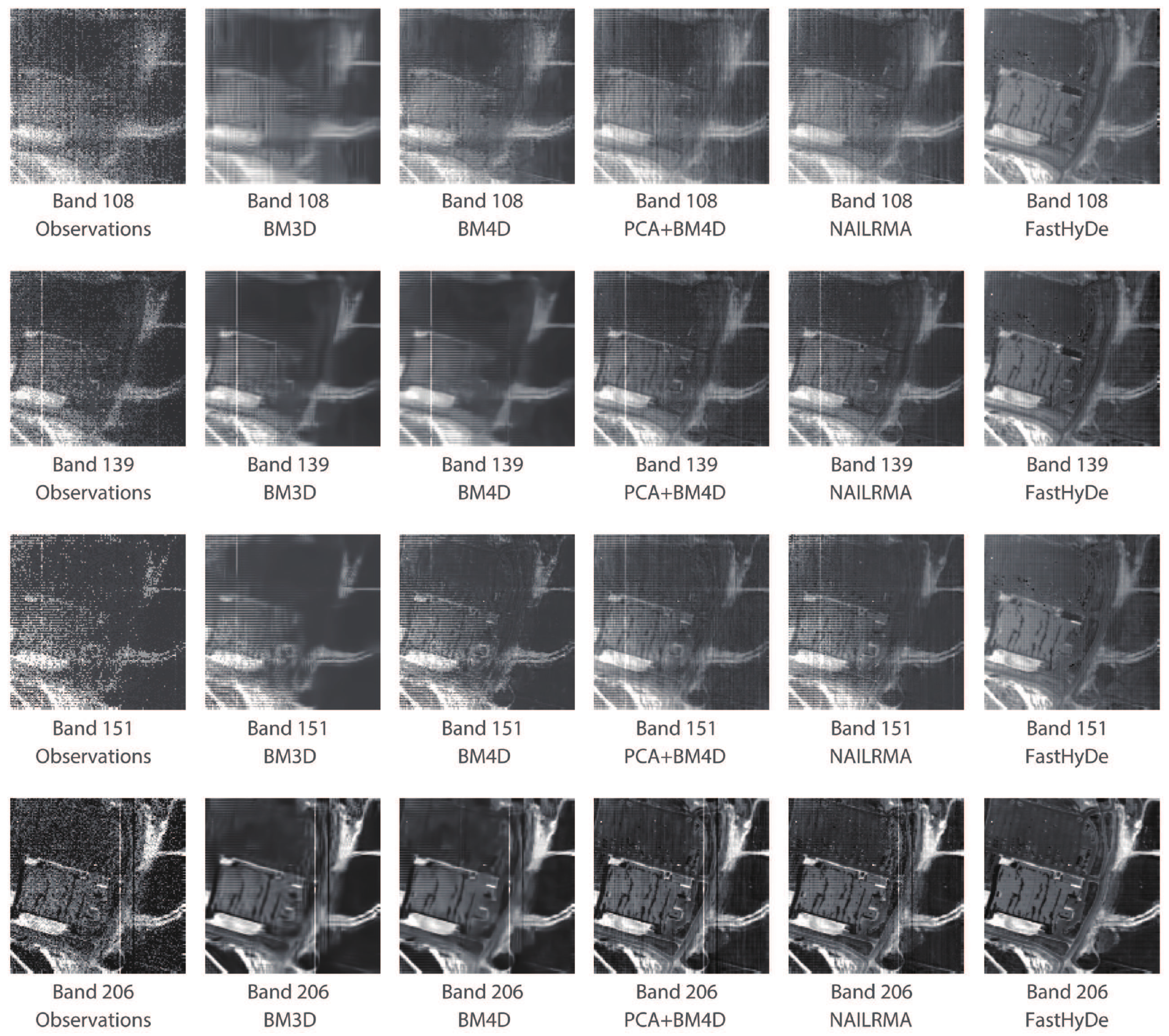}
\caption{Denoising results of BM3D (64 s), BM4D (2579 s), PCA+BM4D (2405 s), NAILRMA (10 s) and FastHyDe (9 s) in Urban subscene.}
\label{fig:Urban_denoised}
\end{figure*}

 \begin{figure*}[htbp]
\centering
\includegraphics[scale=0.8]{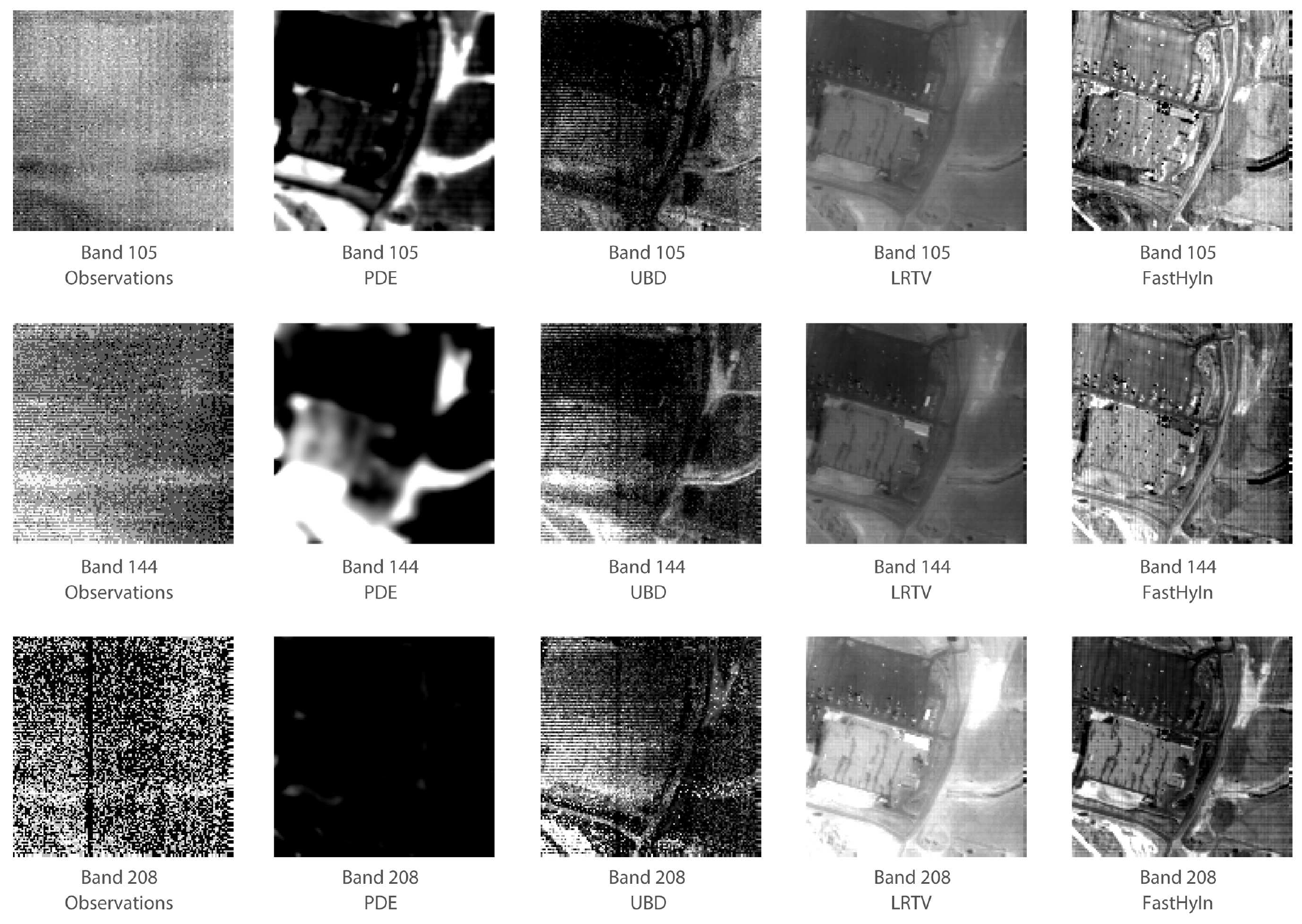}
\caption{ Inpainting results of PDE (86 s), UBD (5 s), LRTV (59 s) and FastHyIn (9 s) in Urban subscene. }
\label{fig:Urban_painted}
\end{figure*}

We apply FastHyIn to Urban subscene (Fig. \ref{fig:RealImg}-(b)), which has $130 \times 130$ pixels, with a spatial resolution of 2 meters per pixel and 210 spectral channels ranging from 400 nm to 2500 nm. 
Due to dense water vapor and atmospheric effects, the image  displays  strong  noise in a number of bands.  
First column in Fig. \ref{fig:Urban_denoised} shows four of those noise bands (208, 139, 151, and 206).

We applied BM3D, BM4D, `PCA+BM4D', NAILRMA and FastHyDe denoisers, under the assumption the noise is  non-i.i.d.
 The parameters used in BM3D and BM4D were set to the same values as in the simulated data. The subspace dimension input to `PCA+BM4D' was 6, which is estimated by HySime \cite{Hysime}. Considering FastHyDe is robust to subspace dimension overestimation, subspace dimension input to FastHyDe was set to 15.
For NAILRMA, the block size and step size were set to 20 and 8, respectively. 
Denoising results are given in Fig. \ref{fig:Urban_denoised}.
Visually, FastHyDe recovers more information in these noisy bands than others do. 
On the other hand, a number of bands in Urban data contain almost no useful information, as shown in first column of Fig. \ref{fig:Urban_painted}. These bands were inpainted by FastHyIn, PDE, UBD,  and LRTV. The parameters of LRTV were manually tuned as $r=6,~\tau = 0.015,~\lambda = 10^5$. The subspace dimension in FastHyIn was set to 15. As shown in \ref{fig:Urban_painted}, the inpainting results of FastHyIn are qualitatively better than those of other inpainters.

\section{Conclusions}

\label{sec:concl}
In this paper, we have proposed a new denoising method for HSIs, termed {\em Fast Hyperspectral denoising} (FastHyDe) and its extended version for inpainting, termed {\em Fast Hyperspectral inpainting} (FastHyIn).  The new methods exploit two characteristics of HSIs: a) they live in low dimensional subspaces, and b) their images of subspace representation coefficients, herein termed 
eigen-images, are self-similar and thus suitable to be denoised with non-local patch-based methods. A comparison of FastHyDe and FastHyIn with the state-of-the-art algorithms is conducted, leading to the conclusion that FastHyDe and FastHyIn  yield  similar or better performance for additive noise and for Poissonian noise,  
with much lower computational complexity.  FastHyDe and FastHyIn are not only fast, but also robust to subspace overestimation, and user-friendly, requiring no parameters, hard to tune.  
These characteristics put FastHyDe and FastHyIn in a privileged position to be used as an  HSI denoiser and inpainter.

\ifCLASSOPTIONcaptionsoff
  \newpage
\fi

\bibliographystyle{IEEEbib}
\bibliography{IEEEabrv,denoisingRef}

\begin{IEEEbiography}
[{\includegraphics[width=1in,height=1.25in,clip,keepaspectratio]{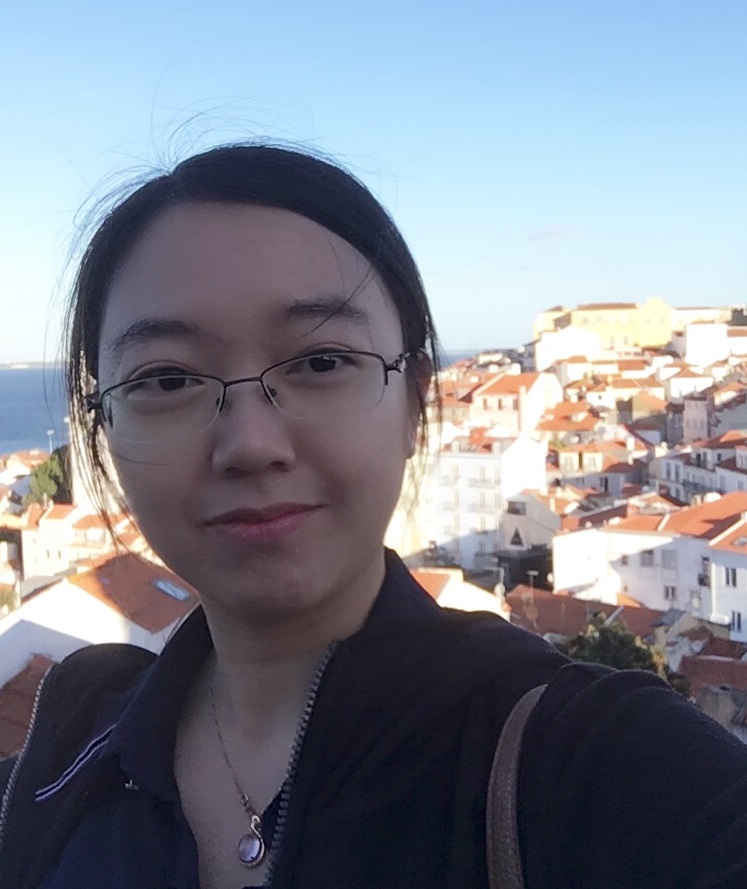}}]{Lina Zhuang}
(S'15) received  Bachelor's  degrees  in geographic  information  system  and  in  economics
from  South  China  Normal  University,  Ghuangzhou,
China, in 2012, and the M.S. degree in cartography and geography information system from Institute of Remote Sensing and Digital
Earth, Chinese Academy of Sciences, Beijing, China, in 2015. She is currently working toward the Ph.D. degree in Electrical and Computer Engineering at  the Instituto Superior T\'{e}cnico, Universidade de Lisboa, Lisbon, Portugal.

Since 2015, she has been with  the Instituto de Telecomunica\c{c}\~{o}es, as   a   Marie   Curie Early Stage Researcher of Sparse Representations and Compressed Sensing Training Network (SpaRTaN number 607290). SpaRTaN Initial Training Networks (ITN) is funded under the European Union's Seventh Framework Programme (FP7-PEOPLE-2013-ITN) call and is part of the Marie Curie Actions--ITN funding scheme.  
Her research interests include hyperspectral image
denoising, inpainting, superresolution, and compressive sensing.
\end{IEEEbiography}

\begin{IEEEbiography}[{\includegraphics[width=1in,height=1.25in,clip,keepaspectratio]{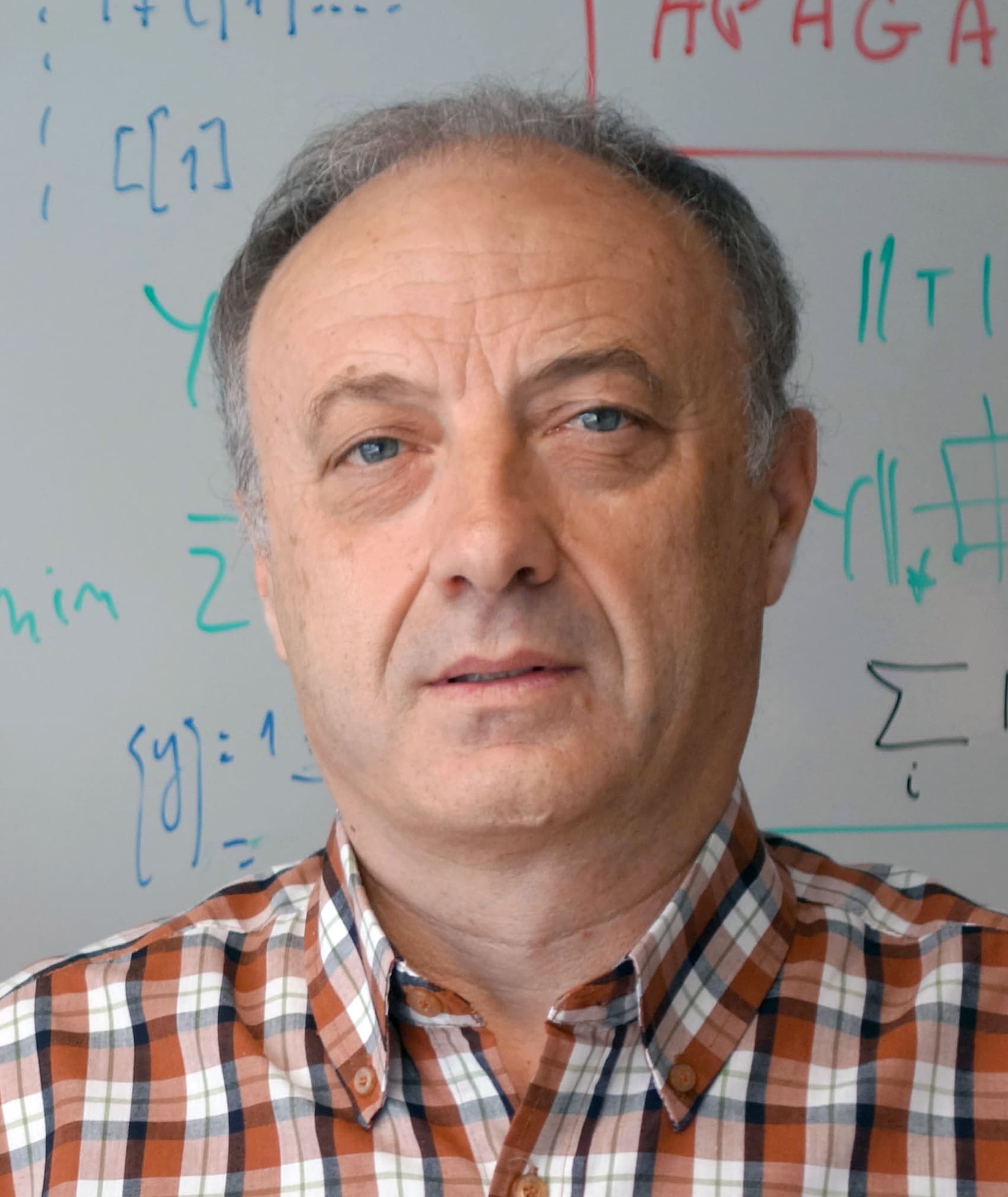}}]{Jos\'{e} M. Bioucas-Dias}
  (S'87--M'95--SM'15--F'17)
received the E.E., M.Sc., Ph.D., and Habilitation degrees in electrical and
computer engineering from Instituto Superior T\'ecnico (IST),
Universidade T\'ecnica de Lisboa (now Universidade de Lisboa),
Portugal, in 1985, 1991, 1995, and 2007, respectively. 

Since 1995, he
has been with the Department of Electrical and Computer Engineering,
IST, where he is currently a Professor and teaches inverse problems
in imaging and electric communications. He is also a Senior Researcher
with the Pattern and Image Analysis group, Instituto de Telecomunica\c{c}\~{o}es, which is a private nonprofit research institution.

His research interests include inverse problems, signal and image
processing, pattern recognition, optimization, and remote sensing.
He has introduced scientific contributions in the areas  of
imaging inverse problems, statistical image processing, optimization,
phase estimation, phase unwrapping, and in various imaging applications,
such as hyperspectral and radar imaging.  

Prof. Bioucas-Dias  was included in Thomson
Reuters' Highly Cited Researchers 2015 list and was the recipient of the
IEEE GRSS David Landgrebe Award for 2017.
\end{IEEEbiography}

% that's all folks
\end{document}